\documentclass[11pt]{article}

\usepackage[numbers,sort&compress,super]{natbib}

\usepackage{graphicx}
\usepackage{amsmath}
\usepackage{authblk}
\usepackage{geometry}
\usepackage{hyperref}
\usepackage{indentfirst}
\usepackage[labelfont={bf}]{caption}
\usepackage{float} 

\makeatletter
\renewcommand{\@seccntformat}[1]{%
  \csname the#1\endcsname.\quad}
\makeatother

\title{A Data-Driven Parametric Reduced-Order Chemical Kinetics Model Derived from Atomistic Simulations}
\author{Michael N. Sakano}
\author{Alejandro Strachan$^*$}

\affil{School of Materials Engineering and Birck Nanotechnology Center, Purdue University, West Lafayette, Indiana, 47907 USA}

\date{}

\begin{document}
\maketitle

\begingroup
\renewcommand\thefootnote{}
\footnotetext{%
\textbf{*Corresponding author:}
\href{mailto:strachan@purdue.edu}{strachan@purdue.edu}%
\par
ORCID MNS: \href{https://orcid.org/0000-0003-3337-4810}{0000-0003-3337-4810}%
}
\endgroup

\begin{abstract}
\noindent
Coarse-grained modeling in molecular simulations serves not only to extend accessible time and length scales beyond atomistic limits, but also to reduce high-dimensional chemical data to low-dimensional representations that expose the underlying latent structure. In the context of energetic materials, reduced-order chemical kinetics models are essential for describing thermally driven decomposition, deflagration, and detonation. Recent data-driven approaches based on machine learning and dimensionality reduction have shown promise for constructing such models directly from atomistic simulations; however, when reaction pathways vary strongly with thermodynamic conditions, these methods can produce latent representations that are difficult to interpret physically or extrapolate reliably. Here, we introduce a parametric, temperature-dependent autoencoder framework that learns a unified reduced-order description of chemical decomposition across a wide range of temperatures within a single model. Physical interpretability is enforced through non-negativity constraints and a softmax activation, enabling the latent variables to be directly associated with additive chemical components and their relative contributions. Reaction kinetics and heat-release parameters are optimized simultaneously within the neural-network architecture, providing a self-consistent coupling between chemical evolution and energetics. The proposed approach yields significantly improved reconstruction accuracy compared to a state-of-the-art dimensionality-reduction method, as quantified by reductions in mean-squared error, while preserving a physically meaningful latent representation. These results demonstrate that parametric, interpretable machine-learning models can provide robust reduced-order chemical kinetics suitable for multiscale modeling of complex reactive systems.

\end{abstract}

\begin{figure}[H]
\centering
\includegraphics[width=\linewidth]{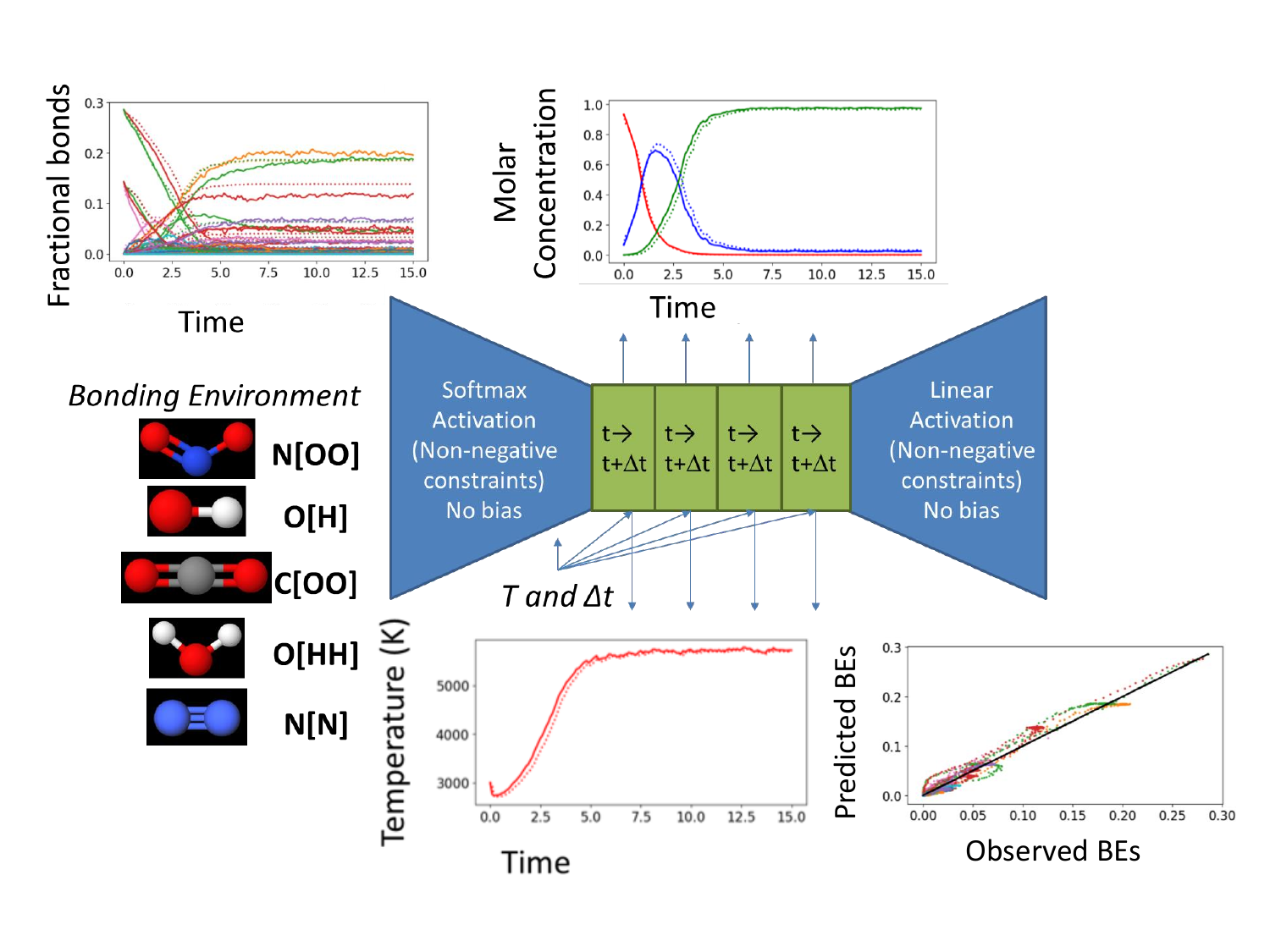}
\captionsetup{labelformat=empty}
\caption*{\textbf{Graphical Abstract}}
\end{figure}

\section{INTRODUCTION}

Predictive modeling of multi-step strongly coupled reactive phenomena, such as shock- and deflagration-driven decomposition in high–energy-density (HED) materials, requires an accurate description of chemical, mechanical, and thermal processes under extreme pressures and temperatures. A fundamental difficulty lies in bridging the wide separation of spatiotemporal scales involved: atomistic bond-breaking and energy redistribution occur on femtosecond–nanosecond and ångström–nanometer scales, whereas macroscopic ignition, deflagration, and detonation evolve over microseconds–milliseconds and micrometers–centimeters. Although atomistic simulations have provided valuable insight into early-stage chemistry and energy localization mechanisms,~\cite{shan2014shock,wood2015ultrafast} their computational cost precludes direct simulation of later-time dynamics. This limitation motivates the development of reduced models that retain essential chemical and thermodynamic fidelity while enabling predictive multiscale descriptions.

A common strategy for achieving this objective is coarse graining, wherein high-dimensional simulation data are compressed into a latent space with substantially fewer degrees of freedom. When applied to reactive systems, coarse graining yields reduced-order chemical kinetics (ROCK) models that approximate complex reaction networks using a small number of effective pathways. In the context of energetic materials, such models have traditionally been constructed by identifying dominant molecular species and fitting global reaction rates to experimental or simulation data. Representative examples for HMX include the single-step Henson–Smilowitz~\cite{henson2002ignition} and Menikoff~\cite{menikoff2006detonation} models, multi-step formulations by Tarver and co-workers,~\cite{tarver1996critical,tarver2004thermal} and more recent extensions incorporating additional intermediates.~\cite{henson2009modeling} While these models reproduce ignition times and one-dimensional cook-off or time-to-explosion measurements, they rely heavily on empirical Arrhenius rate laws and implicit assumptions regarding dominant chemistry. Sensitivity studies have demonstrated that order-of-magnitude variations in rate parameters can qualitatively alter hotspot growth and stability,~\cite{rai2020evaluation} underscoring the lack of consensus regarding an optimal reduced description.

Parallel developments in the broader reactive-simulation community have sought to derive ROCK models directly from molecular dynamics (MD) simulations using reactive force fields such as ReaxFF. Applications span a wide range of systems, including phenolic pyrolysis,~\cite{bauschlicher2013comparison} hydrogen peroxide decomposition,~\cite{ilyin2019first} graphene synthesis,~\cite{lei2019mechanism} and hydrogen combustion.~\cite{wang2011reactive} More recent efforts emphasize automated discovery of reaction pathways and kinetics. Döngen and co-workers introduced algorithms to extract reaction mechanisms and rate constants from ReaxFF simulations of methane oxidation and H$_2$/O$_2$ combustion,~\cite{döntgen2015automated,döntgen2018automated} while Martínez and collaborators developed the ab initio nanoreactor to autonomously explore reaction networks using accelerated first-principles MD.~\cite{wang2014discovering,wang2016automated,pieri2021non,ford2021nitromethane} These approaches represent an important shift toward data-driven chemistry discovery but typically rely on explicit molecular species identification.

Species-based analysis, while natural within ReaxFF and quantum MD frameworks, becomes increasingly challenging as chemical complexity grows. This limitation has motivated alternative representations of chemical evolution based on graph-theoretic concepts. The Simplified Molecular Input Line Entry System (SMILES)~\cite{weininger1988smiles} encodes molecular structure as a graph and has become central to machine-learning applications in chemistry.~\cite{arimoto2005development,worachartcheewan2014large,winter2019learning,elton2019deep} In energetic materials modeling, Lee et al. introduced atomic mass binning to construct ROCK models for RDX cook-off and HMX hotspot deflagration,~\cite{lee2016mirrored,lee2020mirrored} demonstrating good agreement with MD results using a Gibbs free energy–based continuum formulation. Sakano \textit{et al.}~\cite{sakano2021unsupervised} and Kober~\cite{kober2022developing} subsequently introduced bonding environments (BEs), derived from local coordination information in ReaxFF simulations, to characterize evolving chemical structure. This representation is closely related to classical group additivity methods developed by Benson~\cite{cohen1993estimation,constantinou1994new} and enabled construction of a BE-based ROCK model parameterized by heat release.

Related graph-based approaches have been explored in other reactive systems. Dufour-Décieux \textit{et al.} characterized hydrocarbon pyrolysis using local bonding features defined by distance and temporal persistence,~\cite{dufour2021atomic} while Rice \textit{et al.} demonstrated strong sensitivity of predicted kinetics and species lifetimes to bonding definitions in PETN decomposition.~\cite{rice2020heuristics} Graph-theoretic tools such as NetworkX~\cite{hagberg2007exploring} have also been applied to analyze radiation damage in polymers.~\cite{kroonblawd2020quantum} Collectively, these studies highlight both the flexibility of graph-based representations and the sensitivity of reduced models to how molecular connectivity is defined.

Regardless of representation, reactive MD simulations typically generate extremely high-dimensional datasets, motivating further dimensionality reduction. Linear techniques such as principal component analysis (PCA) project data onto orthogonal subspaces that maximize variance,~\cite{halko2011finding,martinsson2011randomized} but the resulting components may include negative contributions and lack direct physical interpretability. Non-negative matrix factorization (NMF), by contrast, yields additive, parts-based representations and has proven advantageous in applications ranging from image analysis to clustering and feature discovery.~\cite{lee1999learning,buciu2004application,zhi2010graph,alexandrov2019nonnegative,nebgen2021neural} More generally, nonlinear dimensionality-reduction methods based on machine learning, particularly autoencoders, provide a flexible framework for learning compact latent representations. Autoencoders consist of an encoder that maps input data to a low-dimensional bottleneck and a decoder that reconstructs the original data; linear autoencoders recover PCA subspaces,~\cite{plaut2018principal} whereas nonlinear architectures capture more complex structure.~\cite{han2018deep,brinker2019deep}

Autoencoders have been extensively applied to molecular dynamics data, especially in biomolecular systems, to identify collective variables, model slow conformational dynamics, and explore free energy landscapes.~\cite{su2022kinetics,brown2008algorithmic,doerr2017dimensionality,chen2018collective,lusch2018deep} Time-lagged and variational formulations have enabled learning of slow modes and propagators,~\cite{wehmeyer2018time,hernandez2018variational} with subsequent extensions improving transferability across systems and force fields.~\cite{sultan2018transferable} Despite these advances, application of autoencoder-based methods to chemically reactive systems remains limited. Reactive dynamics are characterized by rapid, nonlinear changes in bonding environments, posing challenges for standard time-lagged formulations, which tend to perform best when dynamics are smooth or weakly nonlinear. Iterative and stacked architectures have been proposed to mitigate these issues,~\cite{wolpert1992stacked,sridhar1996process} but their utility for learning chemically interpretable reduced models has not been systematically explored.

Our previous work~\cite{sakano2021unsupervised} and those by Kober~\cite{kober2022developing} employed NMF to construct reduced-order chemistry models for ReaxFF simulations conducted under fixed thermodynamic conditions. While this approach revealed clear correlations between latent components and bonding environments, it also highlighted limitations associated with sparse training data and condition-specific models. More broadly, most existing approaches decouple dimensionality reduction from kinetic parameterization, potentially leading to suboptimal or inconsistent reduced descriptions. These observations raise several open questions: Can a unified reduced representation be learned across thermodynamic conditions? Can latent variables and kinetic parameters be optimized self-consistently? And can data-driven methods yield physically interpretable chemical models such as traditional Arrhenius-based formulations?

In this work, we address these challenges by introducing an interpretable parametric autoencoder framework that simultaneously learns reduced chemical representations, reaction kinetics, and heat-release behavior directly from all-atom ReaxFF MD simulations conducted under homogeneous isothermal and adiabatic conditions. We demonstrate that a naïve single-step time-lagged autoencoder fails to recover physically meaningful latent variables due to strong dependence on evolving bonding environments. By initializing the model using condition-specific reduced representations, we recover interpretable collective variables that encode dominant chemical processes. These components are subsequently used to extract reduced-order chemical kinetics and heat-release parameters. Finally, we develop a stacked time-lagged parametric autoencoder trained jointly on isothermal and adiabatic data, enabling self-consistent optimization of structure and kinetics.

The remainder of this paper is organized as follows. Section II describes the simulation data and dimensionality-reduction framework. Section III details the extraction of reduced-order chemical kinetics and heat release. Section IV introduces the stacked parametric autoencoder and its self-consistent optimization. Conclusions and outlook are presented in Section V.

\section{METHODS}

\subsection{Atomistic simulations and molecular feature definition}

In our previous work,~\cite{sakano2021unsupervised} we developed a reduced-order chemical kinetics framework based on reactive molecular dynamics (MD) simulations of homogeneous, isothermal, and adiabatic decomposition of 1,3,5-trinitro-1,3,5-triazinane (RDX). The present study builds upon that dataset and methodology. All simulations were performed using the reactive force field ReaxFF, which describes bond formation and dissociation through a continuous bond-order formalism dependent on interatomic separation. We employ the ReaxFF parameterization developed by Wood \textit{et al.},~\cite{wood2014coupled} which has been shown to reproduce key features of thermally and shock-induced chemistry in high–energy materials.~\cite{islam2019reactive,powell2020insight} Partial atomic charges were updated at every timestep using the charge equilibration (QEq) method.~\cite{aktulga2012parallel} All simulations were carried out using the LAMMPS molecular dynamics package~\cite{plimpton1995fast} (version 16Mar2018).

The subsections below summarize the simulation protocol and introduce the coordination-geometry–based molecular descriptors used for subsequent dimensionality reduction and reduced-order modeling.

\subsubsection{Sample preparation and homogeneous decomposition simulations}

Initial configurations were generated following the procedures described in Refs.~\citenum{sakano2021unsupervised} and ~\citenum{sakano2018role}. A supercell of crystalline RDX at experimental density was constructed by replicating the unit cell~\cite{hakey2008redetermination} in a 3 × 3 × 3 arrangement, followed by energy minimization.

For isothermal simulations, twelve systems with initial temperatures ranging from 1200 to 3000 K were prepared by assigning atomic velocities consistent with the target temperature. Temperature control was maintained using a Nosé–Hoover thermostat in the canonical (NVT) ensemble. Depending on temperature, simulations were run for 100 ps to 1 ns, or until the concentrations of major decomposition products reached steady state.

Adiabatic decomposition simulations were generated by initializing an additional twelve systems at twice the corresponding isothermal temperatures (2400–6000 K) and evolving them in the microcanonical (NVE) ensemble. The integration timestep was chosen to minimize energy drift at elevated temperatures: 0.1 fs for T$_0$$<$1800 K, 0.05 fs for 1800$\le$T$_0$$<$2400 K, and 0.025 fs for T$_0$$\ge$2400 K.

\subsubsection{Coordination geometry analysis}

Thermal decomposition proceeds through a large number of transient intermediates, radicals, and final products. Explicitly tracking all molecular species results in a prohibitively high-dimensional description, while restricting attention to a predefined set of species risks excluding condition-dependent or emergent chemistry. To avoid these limitations, we adopt an unsupervised, data-driven representation based on intrinsic structural features, without imposing \textit{a priori} assumptions about molecular identity or reaction pathways.

Molecular descriptors are defined using local coordination geometry based on first-nearest-neighbor interactions. Specifically, we utilize the ReaxFF bond table output, which identifies atom–atom connectivity using a bond-order cutoff of 0.3. Each atom is characterized by the elemental types of its bonded neighbors, defining a local bonding environment (BE). For the CHNO chemical system considered here, and assuming a maximum coordination number of four, this representation yields a finite set of 280 distinct bonding environments.

\begin{figure}[!hb]
\centering
\includegraphics[width=0.6\linewidth]{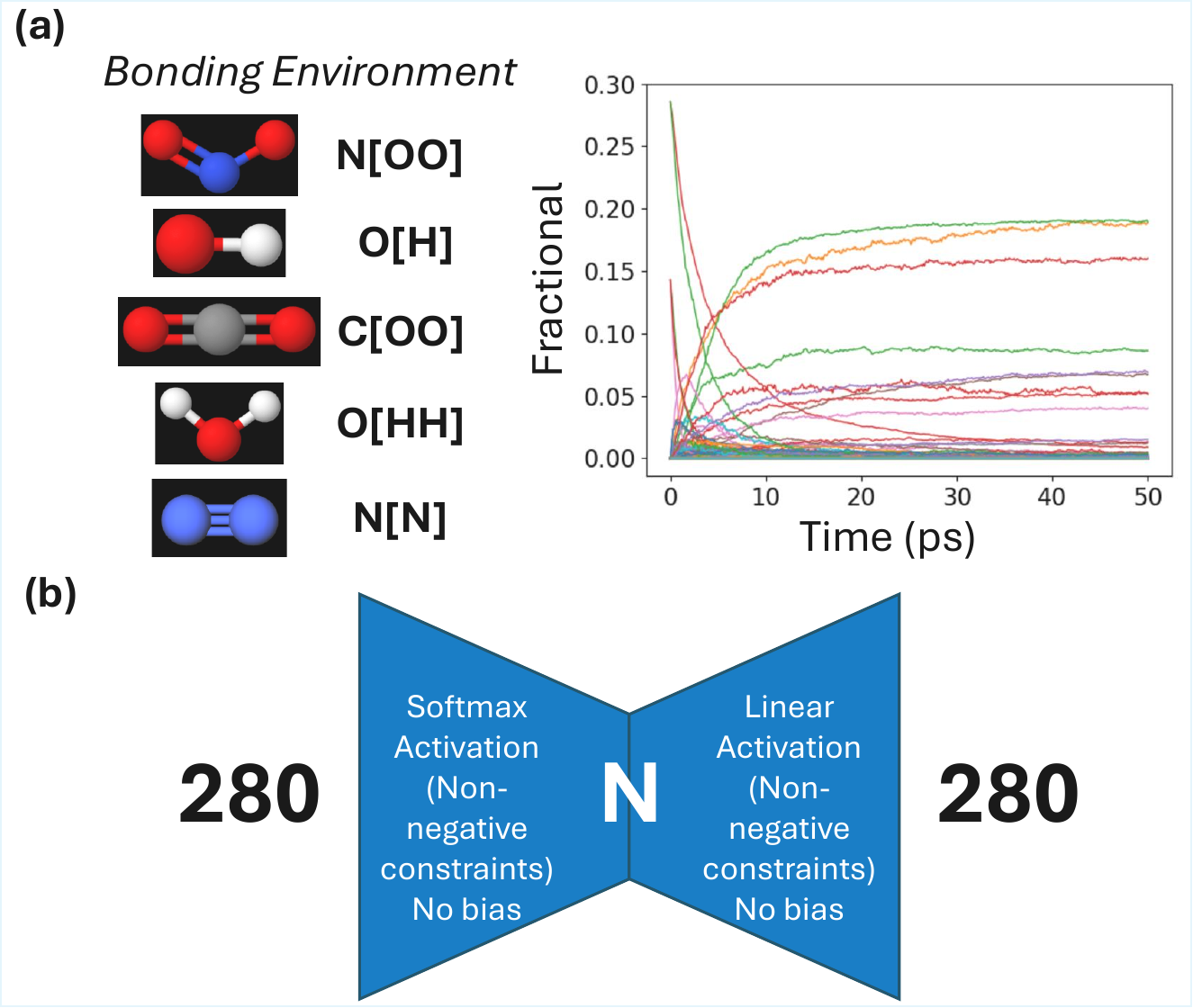}
\caption{\label{Figure1}Representative bonding environments (BEs) and schematic overview of the feedforward autoencoder architecture. (a) Examples of local bonding environments shown using both pictorial representations and corresponding alphabetic notations (left). The associated time series of all 280 BE descriptors obtained from a molecular dynamics simulation are shown on the right. (b) Architecture of the autoencoder, which takes the 280-dimensional BE descriptor vector as input and encodes it into a reduced latent space of dimension \textit{N}. The latent variables are subsequently decoded to reconstruct the original BE populations. Non-negativity constraints and activation functions enforcing unity summation of the encoded variables are applied to promote physically interpretable latent representations, as indicated schematically.}
\end{figure}

For each simulation, time-resolved populations of all bonding environments are computed, resulting in a 280-dimensional descriptor vector at each timestep. Representative bonding environments and their temporal evolution at a selected isothermal condition are shown in Figure~\ref{Figure1}(a). These high-dimensional time series form the input to the dimensionality-reduction procedures described below.

\subsection{Dimensionality reduction architecture and implicit global models}\label{SectionIIB}

The goal of this work is to construct a reduced-order chemistry representation that coarse grains the 280-dimensional bonding-environment (BE) space into a small number of physically interpretable components. In our prior study,~\cite{sakano2021unsupervised} we employed non-negative matrix factorization (NMF)~\cite{cichocki2009fast,fevotte2011algorithms} to achieve this reduction, motivated by the non-negativity of BE populations and their interpretation as additive chemical components. While NMF yielded interpretable latent variables, it proved difficult to construct a single global model that implicitly captured temperature dependence.

Specifically, the BE weights obtained from NMF exhibited systematic but subtle variations with temperature. For example, at higher temperatures the contribution of RDX-like bonding environments to the dominant component decreased, consistent with accelerated dissociation expected from Arrhenius behavior. Although the temperature-specific NMF models were qualitatively similar, their lack of a unified representation precluded extrapolation beyond the training conditions. Attempts to construct a global NMF model using standard implementations in SciPy~\cite{virtanen2020scipy} were unsuccessful.

These limitations motivated the development of a custom dimensionality-reduction framework based on autoencoders. A schematic of the feedforward autoencoder architecture used in this study is shown in Fig.~\ref{Figure1}(b). The model was implemented using the Keras application programming interface~\cite{chollet2017xception} (version 2.2.5) with TensorFlow~\cite{abadi2016tensorflow} (version 1.13.1) as the backend. The autoencoder takes the 280-dimensional BE descriptor vector as both input and output. The autoencoder maps the 280-dimensional BE descriptor vector to a lower-dimensional latent space and reconstructs the original input.

The encoder employs a softmax activation function with non-negative constraints imposed on the kernel weights. This choice enforces interpretability by ensuring that the latent variables are non-negative and normalized, such that their sum at each timestep is unity. Although softmax activations are typically used for classification, here they are used to represent the relative contributions of coarse-grained chemical components. At early times, a component dominated by RDX bonding environments is expected to prevail, whereas at later times components associated with intermediates and products dominate.

No bias term is included in the encoding layer. Preliminary tests indicated that including a bias resulted in negligible improvements in reconstruction accuracy (reductions in mean squared error on the order of 10$^{-8}$ for baseline values of $\sim$10$^{-6}$) and did not qualitatively affect the learned latent components. The bias term was therefore omitted for simplicity.

The latent space dimension was set to \textit{N}=3, consistent with our previous findings~\cite{sakano2021unsupervised} that three components yield chemically interpretable representations corresponding to reactants, intermediates, and products. The decoder reconstructs the BE populations using a linear activation function with non-negative kernel constraints and no bias term. The model was trained using the mean squared error (MSE) as the objective function.

We also explored deeper autoencoder architectures with additional hidden layers and intermediate dimensionalities ranging from 3 to 280. Although deeper networks offer increased representational capacity,~\cite{lipton2015critical,wang2016learning} these models consistently produced latent variables that were difficult to interpret chemically. Investigation of deep architectures that retain physical interpretability is therefore deferred to future work.

To construct an implicit global chemistry model, the autoencoder was first trained using BE data aggregated from all isothermal simulations. Representative encoded trajectories and parity plots for BE reconstruction are shown in Fig. S1 of the Supplementary Material. The resulting latent variables do not exhibit the expected cascade behavior observed in temperature-specific models; instead, a temperature-dependent trade-off emerges between the second and third components. Similar behavior was observed for global NMF models (Fig. S2), indicating that the difficulty arises from the implicit treatment of temperature dependence rather than the choice of dimensionality-reduction method.

\subsection{Temperature-dependent global model}

The results above indicate that temperature-dependent variations in chemical pathways are not readily captured through implicit global dimensionality reduction. We therefore adopt an explicit, parametric strategy in which temperature dependence is incorporated directly into the autoencoder weights.

This section proceeds as follows. In Sec. II C.\ref{SectionIIC1}, we analyze autoencoders trained independently at each isothermal condition and identify systematic temperature-dependent trends in the learned kernel weights. In Sec. II C.\ref{SectionIIC2}, we exploit these trends to construct a unified, parametric autoencoder whose weights are explicit functions of temperature.

\subsubsection{Individual temperature-dependent components}\label{SectionIIC1}

\begin{figure}[H]
\centering
\includegraphics[width=\linewidth]{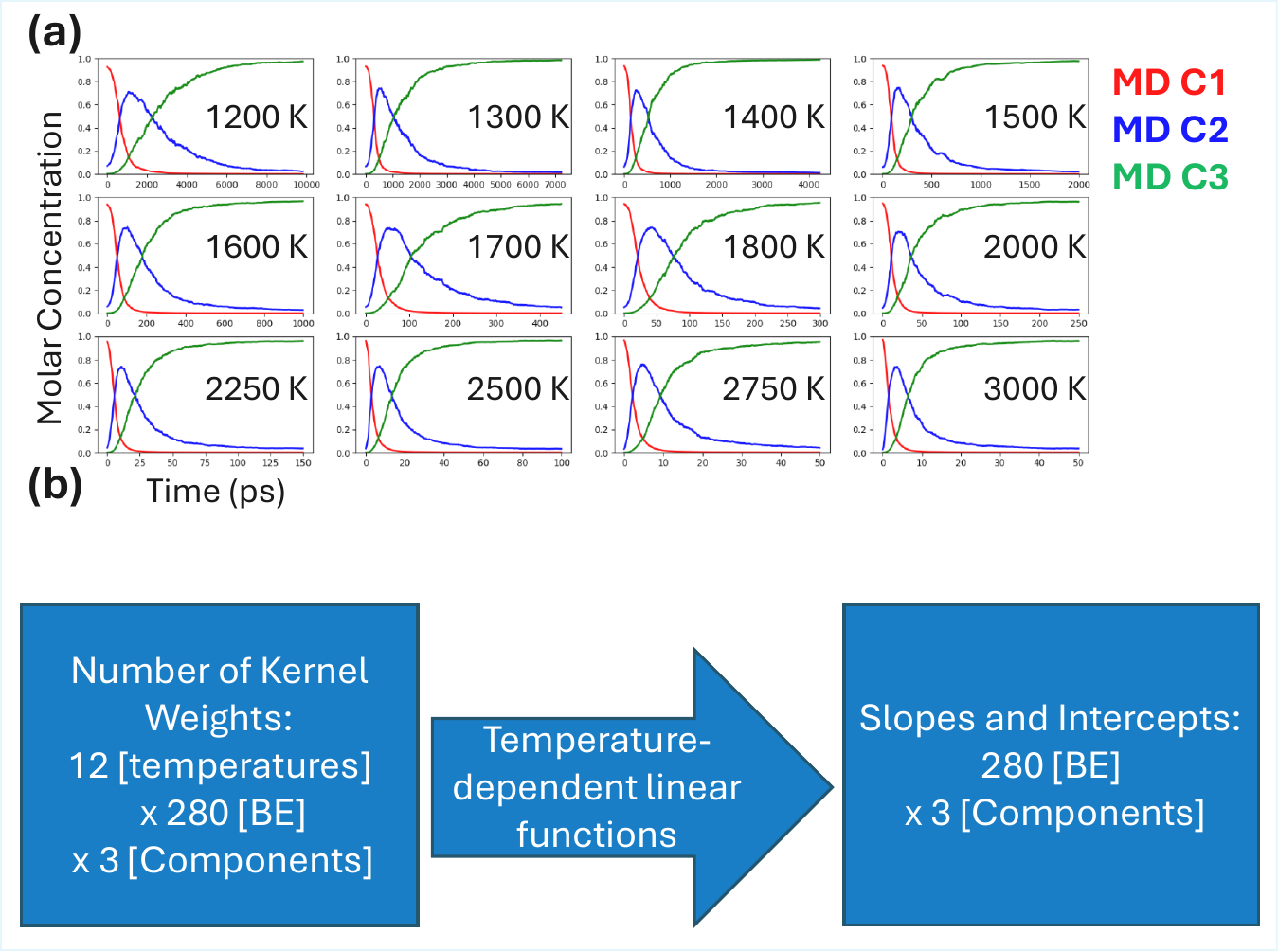}
\caption{\label{Figure2}Temperature-specific autoencoder models and construction of temperature-dependent kernel weights. (a) Encoded latent variables obtained from autoencoder models trained independently at each isothermal temperature, showing the evolution of three interpretable components associated with reactants, intermediates, and products. (b) Schematic illustrating the extraction of temperature-dependent kernel weights: slopes and intercepts are obtained by fitting the encoder and decoder weights from the individual temperature-specific models as functions of isothermal temperature, enabling reconstruction of kernel weight matrices at arbitrary temperatures within the training range.}
\end{figure}

Independent autoencoders were trained for each isothermal temperature using the architecture described in Sec. II \ref{SectionIIB}. Reconstruction errors for these temperature-specific models are uniformly low, with a representative mean squared error of 2.31×10$^{-6}$, as shown in Fig.~\ref{Figure2}(a).

The encoded latent variables exhibit behavior consistent with both physical intuition and our previous NMF-based analysis: an early-time component dominated by intact RDX bonding environments, a transient intermediate component, and a late-time component dominated by stable products. This consistency across modeling approaches supports the physical interpretability of the learned representations.

To assess temperature dependence, we examined the encoder and decoder kernel weights for each temperature-specific model. Representative weights with the largest magnitudes are shown in Fig. S3 of the Supplementary Material as functions of temperature. Bonding environments characteristic of the parent RDX molecule (e.g., C[HHNN], N[CCN], N[NOO]) dominate the first component, whereas environments associated with stable products (e.g., N[N], C[OO], O[HH]) dominate the third. Importantly, the kernel weights vary smoothly with temperature and are well approximated by linear functions over the temperature range considered.

This observation motivates a parametric representation in which each kernel weight is expressed as a linear function of temperature. For each bonding environment and latent component, linear regression is used to extract slopes and intercepts describing this dependence. A schematic of this procedure is shown in Fig.~\ref{Figure2}(b).

\subsubsection{Parametric, interpretable autoencoder components}\label{SectionIIC2}

To construct a unified global model, we define a parametric autoencoder whose encoder and decoder weights are explicit functions of temperature. Using the modular design of Keras, a Python function is implemented that takes temperature as an input parameter and evaluates the corresponding kernel weights from the precomputed linear regressions.

\begin{figure}[H]
\centering
\includegraphics[width=0.8\linewidth]{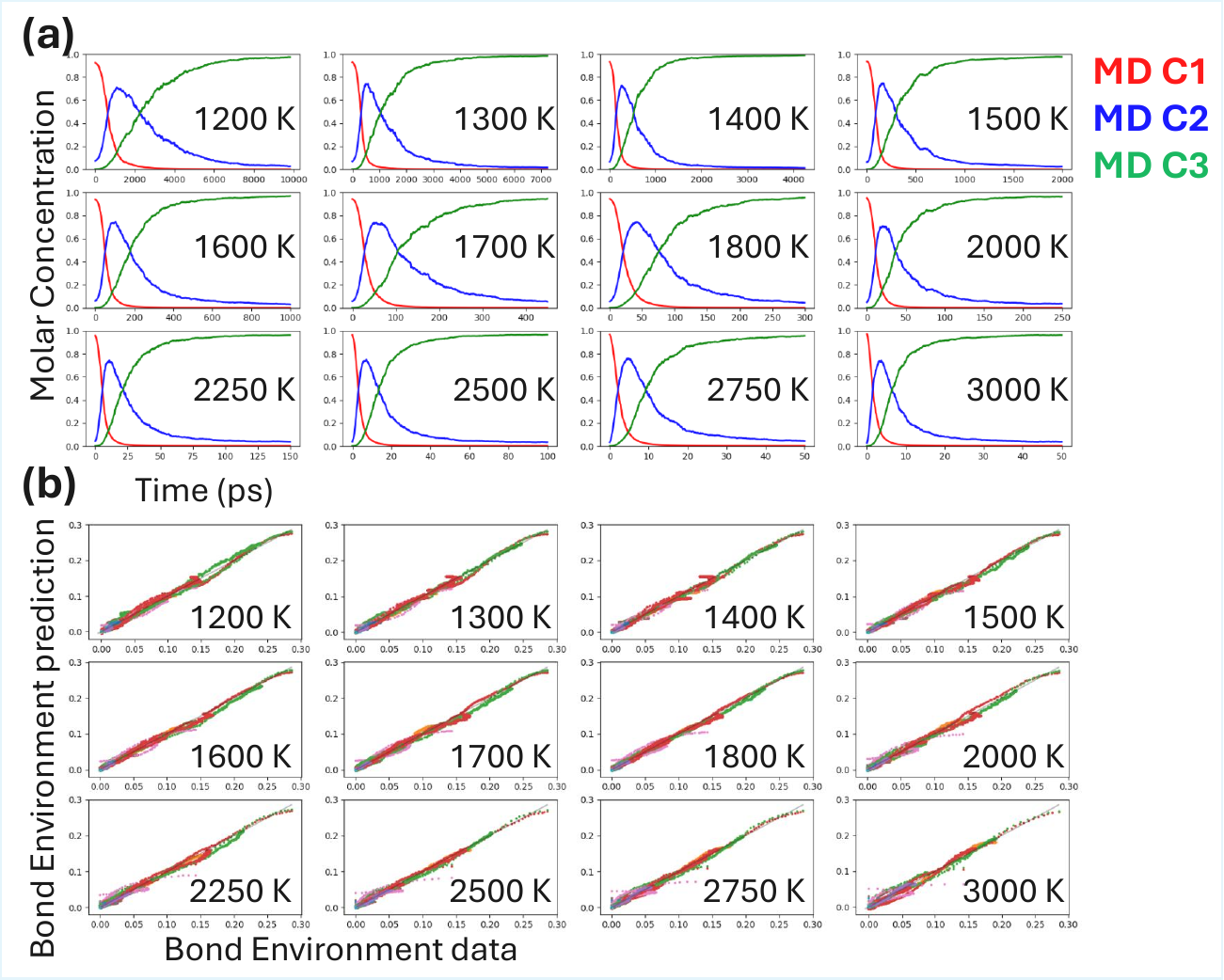}
\caption{\label{Figure3}Performance of the parametric, interpretable autoencoder. (a) Encoded latent variables obtained from the temperature-dependent autoencoder, in which the isothermal temperature is provided as an explicit input to construct the encoder and decoder kernel weights for each component. The resulting latent-space trajectories exhibit consistent, physically interpretable behavior across temperatures. (b) Parity plot comparing reconstructed bonding-environment populations from the autoencoder to the underlying molecular dynamics data, demonstrating high reconstruction accuracy with a mean squared error of 2.31 × 10$^{-6}$ across all temperatures.}
\end{figure}

The resulting autoencoder accepts both the BE descriptor time series and the isothermal temperature as inputs and outputs the corresponding latent-space representation. Figure~\ref{Figure3}(a) illustrates the architecture, and Fig.~\ref{Figure3}(b) compares its predictions with the underlying MD data. The reconstruction accuracy of the parametric model is excellent, with a mean squared error of 2.31×10$^{-6}$, comparable to that of the independently trained temperature-specific models.

For comparison, we implemented an analogous temperature-parameterization strategy using NMF (Fig. S4). While this approach produced slightly lower reconstruction error, the encoded latent variables occasionally assumed negative values and exhibited transient, nonphysical behavior, limiting their interpretability as chemical components. Further analysis indicated that the loss of non-negativity arose from the methodological constraint of combining multiple isothermal temperature weight matrices into a single, global representation, see Supplementary Material Secion S-3 for more details. As a result, this NMF-based parametric model was deemed unsuitable for subsequent kinetic analysis.

In the remainder of this work, we therefore adopt the parametric, interpretable autoencoder and interpret its latent variables as coarse-grained reactant, intermediate, and product components. In Sec. III, we introduce a reduced reaction scheme linking these components, extract temperature-dependent kinetic and heat-release parameters, and validate the resulting reduced-order chemistry model against both isothermal and adiabatic molecular dynamics simulations.

\section{REDUCED-ORDER CHEMICAL KINETICS AND HEAT RELEASE OPTIMIZATION}

We adopt the reduced-order chemical kinetics (ROCK) reaction scheme introduced in our previous work.~\cite{sakano2021unsupervised} Under isothermal conditions, the temporal evolution of the three coarse-grained chemical components is described by the following system of ordinary differential equations:

\begin{align}
\dot{C}_1 &= -C_1 Z_1 \exp\!\Bigg(-\frac{E_1}{R T}\Bigg), \label{Equation1} \\
\dot{C}_2 &= C_1 Z_1 \exp\!\Bigg(-\frac{E_1}{R T}\Bigg) - C_2 Z_2 \exp\!\Bigg(-\frac{E_2}{R T}\Bigg), \label{Equation2} \\
\dot{C}_3 &= C_2 Z_2 \exp\!\Bigg(-\frac{E_2}{R T}\Bigg), \label{Equation3}
\end{align}

where $C_1$, $C_2$, and $C_3$ denote the molar concentrations of the reactant-, intermediate-, and product-like components obtained from the parametric autoencoder. The parameters $Z_1$ and $Z_2$ are pre-exponential factors, $E_1$ and $E_2$ are activation energies, $R$ is the universal gas constant, and $T$ is the temperature.

For adiabatic conditions, chemical kinetics are coupled to thermal evolution through the energy balance

\begin{equation}
    \rho C_v \dot{T} = -Q_1 \dot{C}_1 + Q_2 \dot{C}_2
\label{Equation4}
\end{equation}

where $\rho$ is the system density, $C_v$ is the classical specific heat of RDX, and $Q_1$ and $Q_2$ represent the heat-release parameters associated with the first and second reaction steps, respectively.

In Sec. III~\ref{SectionIIIA}, we parameterize Eqs.~\ref{Equation1}–~\ref{Equation3} using the latent variables generated by the parametric autoencoder and isothermal MD data. In Sec. III~\ref{SectionIIIB}, Eq.~\ref{Equation4} is used to extract heat-release parameters from adiabatic simulations following the segmented regression approach described in Ref.~\citenum{sakano2021unsupervised}.

\subsection{Parametric autoencoder coupled with physics-based kinetics}\label{SectionIIIA}

The interpretable latent variables produced by the parametric global autoencoder are treated as coarse-grained concentration profiles and fitted to Eqs.~\ref{Equation1}–~\ref{Equation3}. Time integration is performed using a forward Euler scheme. For each isothermal temperature, trial kinetic parameters are proposed, the corresponding concentration trajectories are computed, and the resulting profiles are compared directly to the encoded autoencoder components. An optimizer minimizes the mean squared error (MSE) between the ROCK predictions and the latent-space trajectories across all temperatures.

\begin{figure}[H]
\centering
\includegraphics[width=\linewidth]{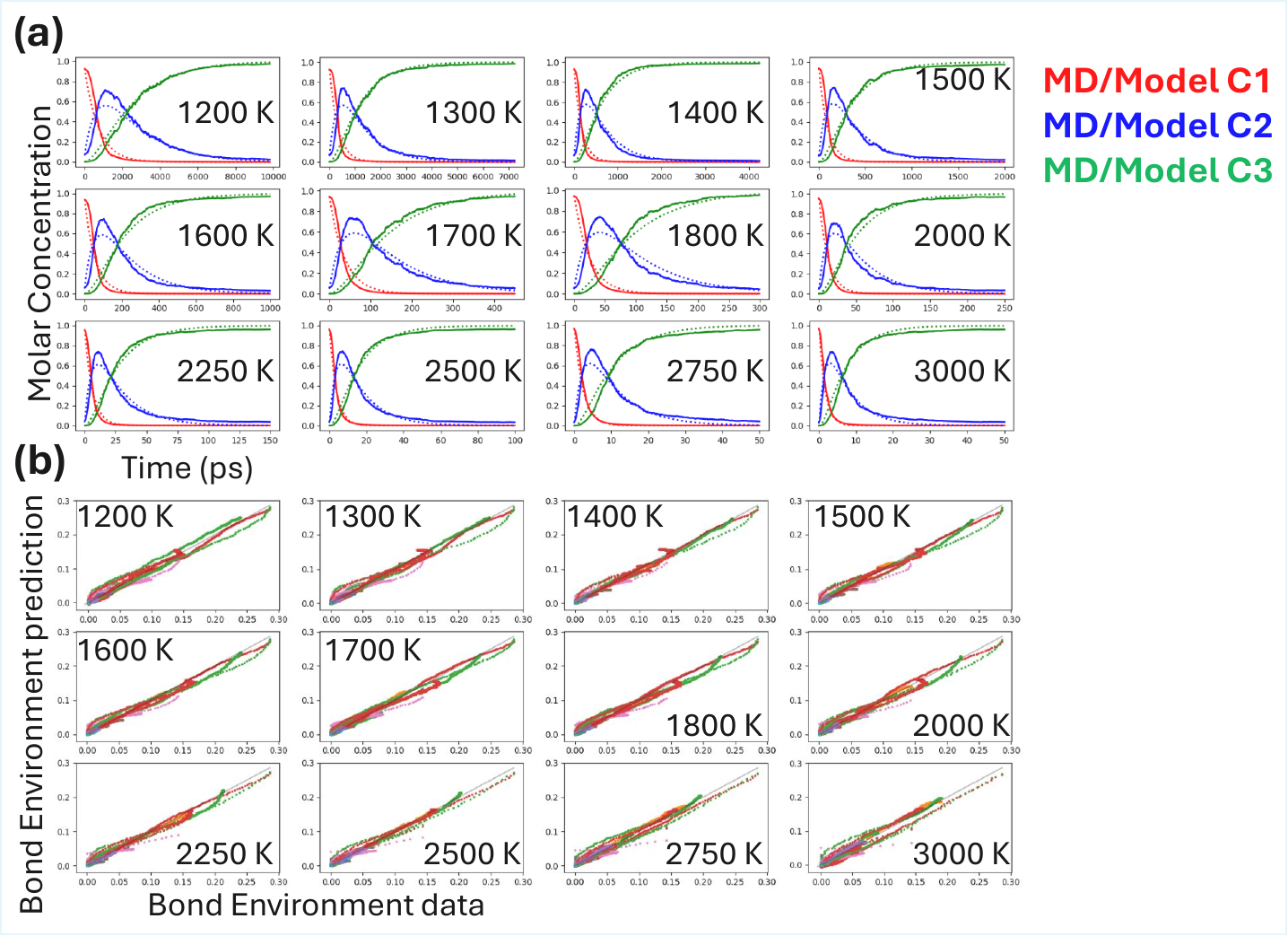}
\caption{\label{Figure4}Parametrized reduced-order chemical kinetics model derived from the interpretable global autoencoder. (a) Comparison between the encoded latent variables obtained from the autoencoder (solid lines) and the corresponding predictions of the physics-based ROCK model (dotted lines) across all temperatures. The mean squared error over all latent components is 1.92 × 10$^{-3}$. (b) Parity plot comparing bonding-environment populations reconstructed by decoding the ROCK model predictions with those obtained directly from the molecular dynamics simulations. The low mean squared error of 3.27 × 10$^{-6}$, evaluated over all bonding environments and temperatures, demonstrates the high fidelity of the parametrized kinetics model.}
\end{figure}

Figure~\ref{Figure4}(a) compares the fitted ROCK model predictions (dashed lines) with the autoencoder-encoded components (solid lines). The optimized kinetic parameters are

\begin{center}
\begin{minipage}{0.45\textwidth}
\centering
$Z_1 = 35.481~\mathrm{ps}^{-1},\quad E_1 = 24.301~\mathrm{kcal\,mol}^{-1}$
\end{minipage}%
\hfill
\begin{minipage}{0.45\textwidth}
\centering
$Z_2 = 7.693~\mathrm{ps}^{-1},\quad E_2 = 23.000~\mathrm{kcal\,mol}^{-1}$
\end{minipage}
\end{center}

Relative to Ref.~\citenum{sakano2021unsupervised}, the prefactor $Z_1$ is modestly larger, while $Z_2$ is approximately four times smaller. Because the activation energies remain comparable, these differences cannot be attributed to compensating prefactor–activation energy correlations. Instead, they reflect differences in the representation of bonding-environment populations between the autoencoder-based and NMF-based encodings.

The resulting MSE for the concentration profiles across all temperatures is 1.92 × 10$^{-3}$. When the fitted concentrations are decoded back into bonding-environment (BE) space using the autoencoder decoder, the reconstructed BE populations show excellent agreement with the underlying MD data, with an overall MSE of 3.27 × 10$^{-6}$. Representative parity plots are shown in Fig.~\ref{Figure4}(b).

For comparison, the same kinetic fitting procedure was applied to both the global and temperature-dependent NMF models developed previously. The resulting concentration trajectories and BE parity plots are shown in Figs. S5 and S6 of the Supplementary Material. In both cases, the NMF-based models exhibit larger errors in both latent-space concentrations and reconstructed BEs than the parametric autoencoder framework, underscoring the advantage of combining temperature-parametric dimensionality reduction with physics-based kinetics.

\subsection{Coupling kinetics with heat release under adiabatic conditions}\label{SectionIIIB}

\begin{figure}[htbp]
\centering
\captionsetup{type=table} 
\caption{\label{Table1}Heat-release parameters obtained by fitting the physics-based reduced-order chemical kinetics model to adiabatic molecular dynamics simulation data using the parametric, interpretable autoencoder framework.}
\includegraphics[width=0.8\linewidth]{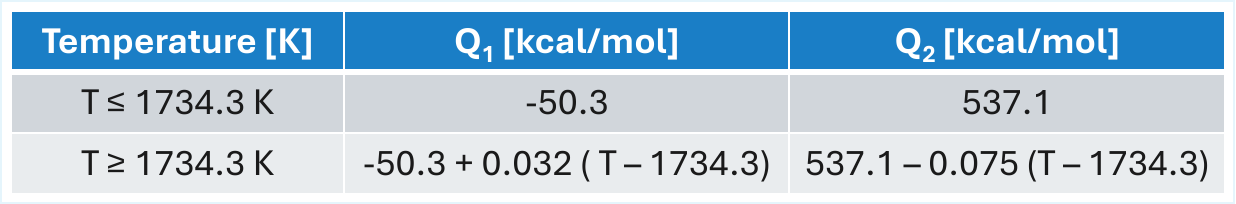}
\end{figure}

With kinetic parameters fixed from the isothermal analysis, we next incorporate heat release to model adiabatic decomposition. Equation~\ref{Equation4} is integrated simultaneously with Eqs.~\ref{Equation1}-~\ref{Equation3} to predict temperature evolution during adiabatic MD simulations. As in Ref.~\citenum{sakano2021unsupervised}, the heat-release parameters $Q_1$ and $Q_2$ are modeled as functions of initial temperature using segmented linear regression. The resulting parameters are summarized in Table~\ref{Table1}.

Relative to Ref.~\citenum{sakano2021unsupervised}, the fitted value of $Q_1$ is approximately 25 times more negative, while $Q_2$ is larger by approximately 25\%. The temperature-dependent slope associated with $Q_2$ is roughly twice as large, whereas both the threshold temperature and slope for $Q_1$ remain consistent between the two studies. These differences reflect the modified partitioning of chemical components resulting from the autoencoder-based representation.

\begin{figure}[H]
\centering
\includegraphics[width=\linewidth]{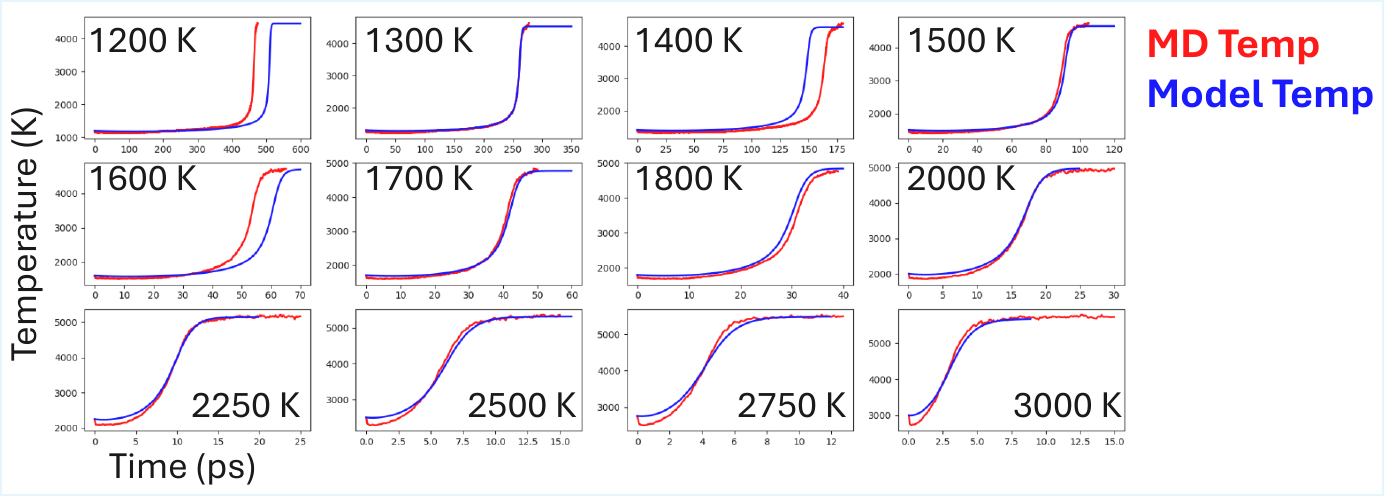}
\caption{\label{Figure5}Adiabatic temperature evolution comparing molecular dynamics simulations with predictions from the reduced-order chemical kinetics model coupled with heat release. Across all initial temperatures, the model accurately captures both the transient temperature evolution and the overall exothermicity, demonstrating the suitability of the chosen functional form for the heat-release parameters.}
\end{figure}

Figure~\ref{Figure5} compares the predicted temperature evolution from the ROCK-plus-heat-release model with the corresponding adiabatic MD data. The model accurately captures both the transient temperature rise and the final equilibrium temperature. Using the mean absolute error (MAE) as the objective function, an overall MAE of 10.8 K is obtained across all training temperatures.

To assess model generalizability, we validate the framework against an adiabatic simulation at an initial temperature $T_0$=1900 K that was not included in the training set. The results, shown in Fig. S7 of the Supplementary Material, demonstrate excellent agreement between MD simulations and model predictions for concentration profiles, temperature evolution, and decoded BE populations. Such validation was not achievable in our earlier NMF-based approach, which relied on temperature-local dimensionality-reduction models.

The ability to accurately reconstruct bonding-environment populations, coarse-grained concentrations, and temperature evolution for previously unseen conditions highlights the utility of the parametric, interpretable autoencoder as a unifying bridge between atomistic simulations and continuum-scale reduced-order chemistry models. In the following section, we extend this framework by introducing a stacked, time-lagged autoencoder architecture that enables self-consistent optimization of kinetic and thermal parameters within a single learning framework.

\section{STACKED TIME-LAGGED PARAMETRIC AUTOENCODER}

\begin{figure}[H]
\centering
\includegraphics[width=\linewidth]{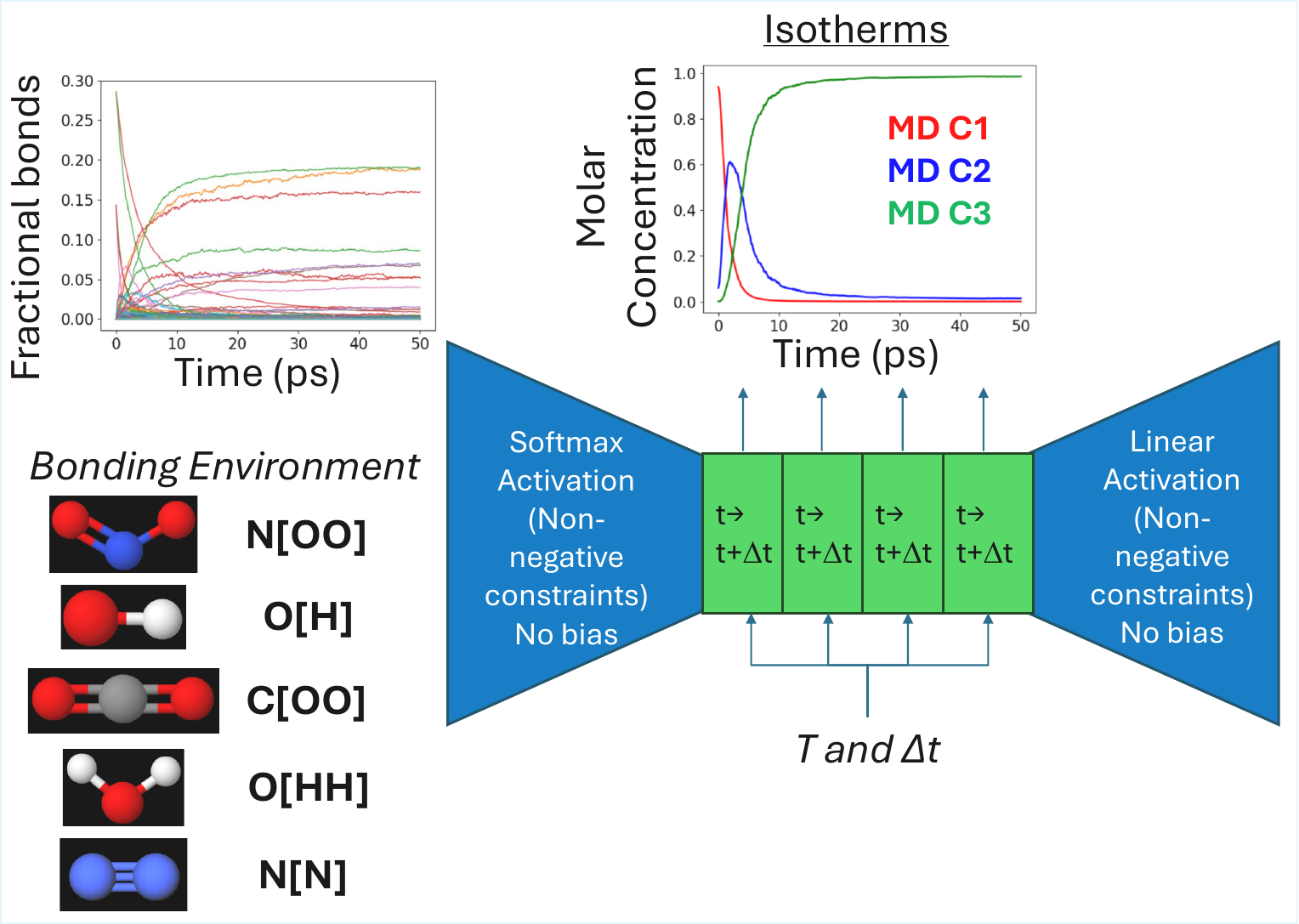}
\caption{\label{Figure6}Time-lagged, stacked parametric autoencoder for isothermal dynamics. The model takes the component concentrations at time $t$, the isothermal temperature $T$, and the timestep $\Delta t$ as inputs and iteratively propagates the system forward to predict concentrations at $t+\Delta t$, where $N$ denotes the number of stacked prediction steps. Multi-step temporal evolution is achieved through repeated application of a single-step propagation network.}
\end{figure}

A schematic of the time-lagged, stacked parametric autoencoder implemented in Keras is shown in Fig.~\ref{Figure6}. The framework operates on the encoded chemical components obtained from the parametric autoencoder and predicts their temporal evolution using a physics-informed propagation network. At each timestep $t$, the network takes as input the component concentrations, the isothermal temperature $T$, and the timestep $\Delta t$, and produces a one-step prediction of the concentrations at $t+\Delta t$. The internal structure of the propagation network, which implements the reduced-order kinetics defined in Sec. III, is shown in Fig. S8 of the Supplementary Material.

To extend this formulation beyond single-step prediction, the one-step propagation network is iteratively applied over multiple timesteps using a stacking procedure. The resulting architecture, illustrated by the green blocks in Fig.~\ref{Figure6}, forms a recurrent prediction framework in which the output concentrations from one step are fed back as inputs to the next. This construction enables explicit examination of error accumulation and long-time predictive behavior. The training data consist of the component concentrations at an initial time $t_0$, the corresponding isothermal temperature, and the timestep, with the model trained to predict concentrations at $t_0+n\Delta t$ for multiple values of $n$. The objective function is defined as the mean squared error (MSE) over all predicted timesteps.

\begin{figure}[H]
\centering
\includegraphics[width=\linewidth]{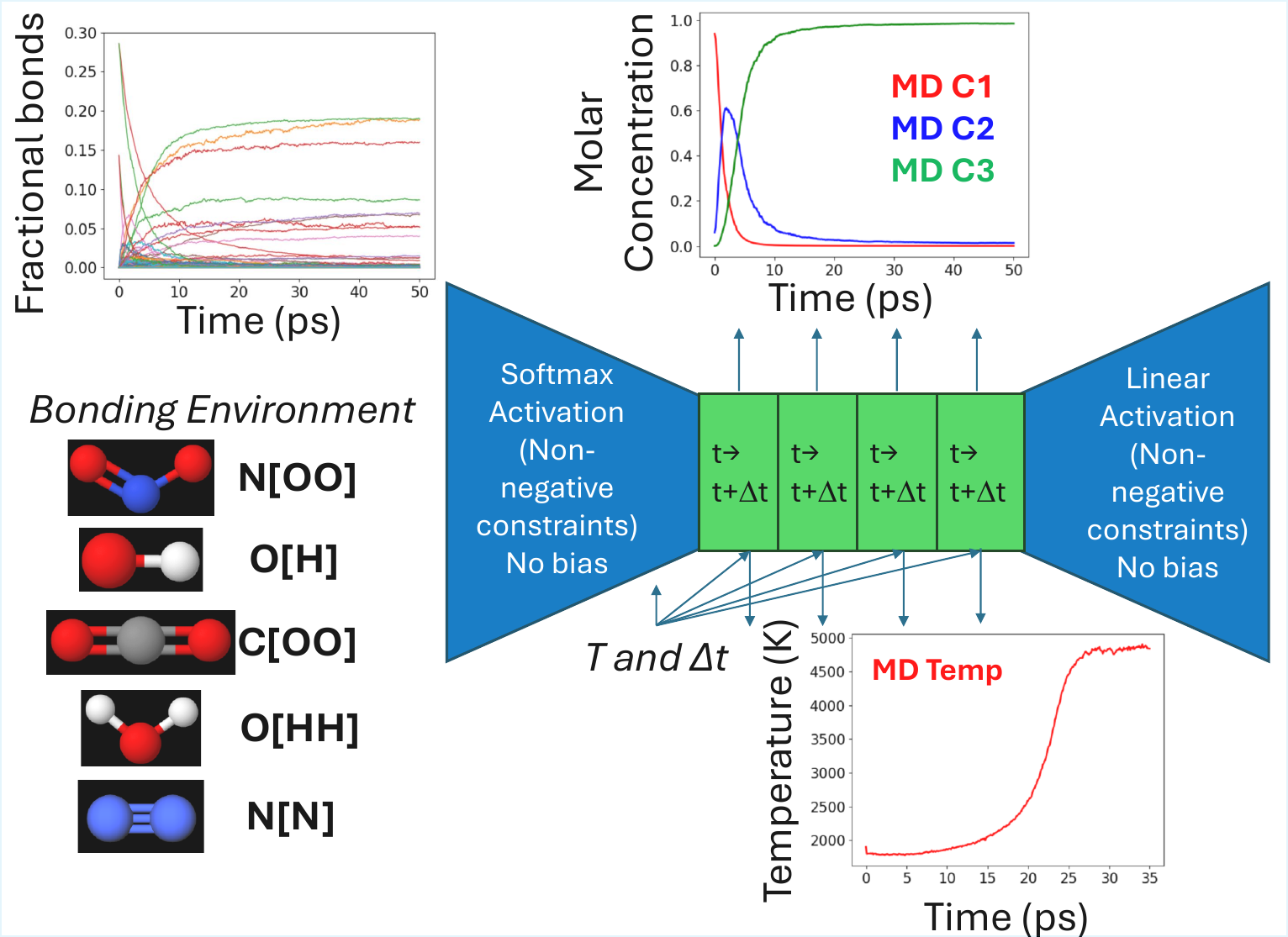}
\caption{\label{Figure7}Time-lagged, stacked parametric autoencoder for adiabatic dynamics. The model takes the component concentrations at time $t$, the temperature $T$, and the timestep $\Delta t$ as inputs and iteratively propagates both concentrations and temperature to $t+n\Delta t$, where $N$ is the number of stacked prediction steps. Adiabatic temperature evolution is incorporated through the coupled kinetics–heat-release formulation.}
\end{figure}

As shown in Fig. S9 of the Supplementary Material, stacked training does not lead to a reduction in the overall MSE relative to the single-step optimization described in Sec. III. The optimized kinetic parameters are identical to those obtained from the single-step fitting procedure, and both the latent-space concentration errors and the decoded bond-environment reconstruction errors remain unchanged. These results indicate that, for the present system and reaction scheme, single-step fitting sufficiently constrains the reduced-order kinetics and that additional temporal stacking does not provide further improvement in predictive accuracy.
We next extend the time-lagged framework to adiabatic conditions by incorporating temperature evolution into the propagation network. A schematic of the inner network used to predict both component concentrations and temperature at the next timestep is shown in Fig. S10 of the Supplementary Material. The inclusion of temperature-dependent heat-release behavior, including switching between reaction stages, required careful construction of the network to ensure numerical stability and physical consistency. Once established, this coupled concentration–temperature predictor was stacked in the same manner as the isothermal model, yielding the recurrent adiabatic architecture shown in Fig.~\ref{Figure7}.

\begin{figure}[H]
\centering
\includegraphics[width=\linewidth]{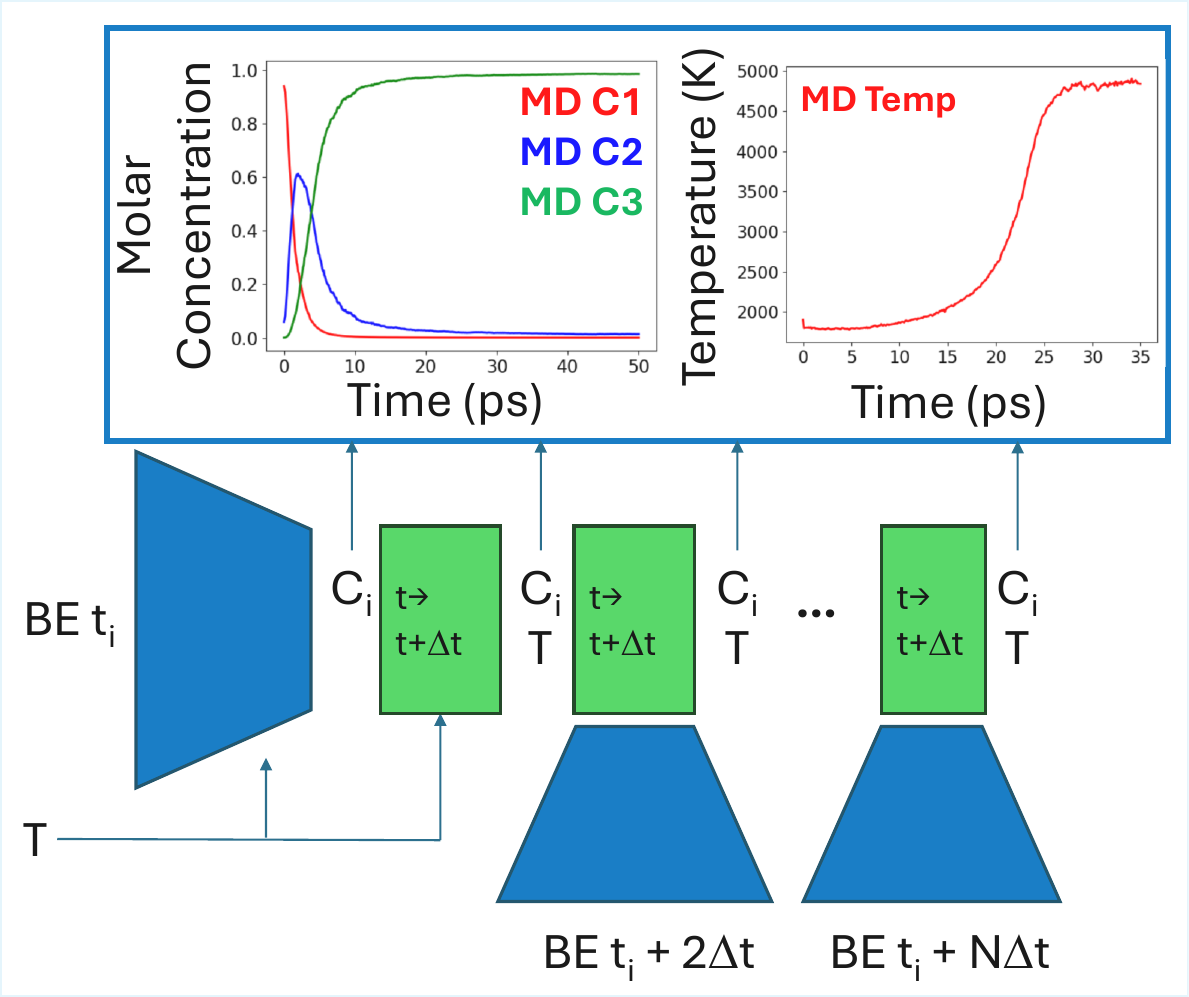}
\caption{\label{Figure8}Self-consistent time-lagged, stacked parametric autoencoder. The architecture follows the construction shown in Figure~\ref{Figure7}, but additionally outputs the decoded bond-environment concentrations at each propagation step. In contrast to the fixed-weight models, the encoder and decoder weights of the parametric, interpretable autoencoder are treated as trainable parameters. This formulation enables simultaneous optimization of the dimensionality reduction and dynamical propagation, and is expected to yield the most accurate and self-consistent reduced-order representation.}
\end{figure}

Finally, we consider a fully self-consistent learning framework in which the parametric autoencoder and the reduced-order chemical kinetics model are optimized simultaneously. A schematic of this integrated approach is shown in Fig.~\ref{Figure8}. In addition to the stacked adiabatic propagation network, the decoded bond-environment populations are evaluated at each timestep and directly compared with the corresponding MD data during training. In this formulation, the encoder and decoder weights of the parametric autoencoder are treated as trainable parameters and optimized concurrently with the kinetic and heat-release parameters.

Preliminary results indicate that joint optimization of the encoder–decoder weights and the dynamical parameters presents significant challenges. In particular, the reconstruction loss associated with the high-dimensional bond-environment populations dominates the overall objective function, effectively suppressing the contribution of the kinetic and thermal evolution constraints. As a result, optimization favors improved reconstruction accuracy at the expense of learning physically meaningful kinetic and heat-release parameters. Unfortunately, multi-step propagation leads to systematic error accumulation that destabilizes the learning process. As prediction errors compound over successive timesteps, the encoder and decoder weights respond by amplifying small discrepancies in the latent representation rather than correcting them, resulting in unbounded growth of the autoencoder parameters. This behavior reflects a mismatch between the objectives of accurate long-horizon reconstruction and stable latent dynamics. Notably, this instability persists even when the inner kinetic and heat-release parameters are fixed, such that the propagation network remains static. In this case, the autoencoder alone attempts to compensate for accumulated dynamical error, but lacks sufficient structural constraints to do so in a physically meaningful manner. As a result, the learned encoder–decoder mappings drift away from the interpretable regime, ultimately leading to numerical divergence, and resulting in failed models.

Addressing this imbalance will require more careful treatment of the multi-term loss function, potentially through normalization, adaptive weighting, or multi-objective optimization strategies that explicitly enforce the relative importance of dynamical fidelity. Although a complete resolution is beyond the scope of the present work, these observations delineate a clear pathway for future development of end-to-end, physically interpretable learning frameworks for reactive molecular systems.

\section{CONCLUSION}

In this work, we have introduced a parametric and interpretable autoencoder framework for coarse graining high-dimensional molecular dynamics data across broad thermodynamic regimes. By explicitly incorporating temperature dependence into the encoder–decoder architecture, the method yields a unified reduced-order representation capable of capturing decomposition chemistry over multiple isothermal conditions within a single model. The resulting latent variables are constrained to be non-negative and normalized, enabling a direct physical interpretation in terms of effective concentrations of reactants, intermediates, and products.

Using this reduced representation, we parameterized reduced-order chemical kinetics (ROCK) models implemented both as standalone numerical solvers and as neural-network-based realizations. The extracted kinetic parameters accurately reproduce the temperature-dependent evolution of the latent components. When coupled to a heat-release model, the resulting framework quantitatively predicts adiabatic temperature evolution, with good agreement relative to molecular dynamics simulations for both training and extrapolative conditions. A key advantage of the parametric autoencoder is its ability to decode bond-environment populations directly from the latent variables, thereby maintaining an explicit connection between coarse-grained dynamics and molecular-scale descriptors, an ability not afforded by earlier condition-specific dimensionality-reduction approaches.

We further extended the framework to time-lagged and stacked parametric autoencoders for modeling both isothermal and adiabatic dynamics. These architectures enable multi-step propagation of component concentrations, and temperature in the adiabatic case, allowing assessment of long-time predictive behavior and error accumulation. Building on this formulation, we outlined a self-consistent optimization strategy in which the reduced representation, reaction kinetics, and heat-release parameters are learned simultaneously, with encoder and decoder weights treated as trainable degrees of freedom. This approach provides a systematic pathway toward end-to-end learning of physically interpretable coarse-grained dynamics from atomistic simulation data.

More broadly, the present results demonstrate that neural-network-based models, when designed with appropriate physical constraints and interpretability requirements, can provide robust and transparent alternatives to traditional coarse-graining and reduced-order modeling techniques. Although the present study focuses on chemical decomposition and reduced-order kinetics, the proposed framework is general and applicable to a wide class of reactive and nonreactive systems characterized by high-dimensional state spaces. Looking forward, parametric and interpretable autoencoder formulations of this kind may facilitate the discovery of effective governing equations directly from data, offering a complementary route to classical theoretical modeling while preserving physical meaning.

\section*{ACKNOWLEDGMENTS}

This work was support by the US Office of Naval Research, Multidisciplinary University Research Initiatives (MURI) Program, Contract N00014-16-1-2557, program managers Clifford Bedford, Chad Stoltz, and Kenny Lipkowitz. The authors would like to thank Saaketh Desai for helpful discussions on Keras implementation.

\section*{AUTHOR DECLARATIONS}

\subsection*{Conflict of Interest}

The authors have no conflicts to disclose.

\subsection*{Author Contributions}

\textbf{Michael N. Sakano:} Conceptualization, Formal Analysis, Investigation, Methodology, Software, Visualization, Writing – Original Draft. \textbf{Alejandro Strachan:} Conceptualization, Data Curation, Methodology, Resources, Supervision, Writing – Review \& Editing.

\section*{DATA AVAILABILITY}

The data that support the findings of this study are available from the corresponding author upon reasonable request.

\section*{SUPPLEMENTARY MATERIAL}

See the supplementary material for more details regarding (i) global models fit to all temperatures simultaneously, (ii) temperature-dependent weights, (iii) the Non-negative Matrix Factorization temperature-dependent model, (iv) kinetics fit to isothermal non-negative matrix factorization global model, (v) predicting the time series of a new adiabatic temperature, and (vi) training the reduced-order chemical kinetics model and heat evolved using Keras.

\renewcommand{\refname}{REFERENCES}

\bibliographystyle{aipnum4-1}
\bibliography{references}

\end{document}


\pagenumbering{arabic}         
\setcounter{page}{1}           
\renewcommand{\thepage}{S\arabic{page}}  

\pagestyle{fancy}
\fancyhf{}                     
\fancyfoot[C]{\thepage}        

\thispagestyle{firstpage}  

\Large \textbf{Supplementary Materials for: A Data-Driven Parametric Reduced-Order Chemical Kinetics Model Derived from Atomistic Simulations}\\[0.3cm]
\large Authors: Michael N. Sakano and Alejandro Strachan$^*$\\
\large Affiliations: School of Materials Engineering and Birck Nanotechnology Center, Purdue University, 
West Lafayette, Indiana, 47907 USA

\section{Global models trained across all temperatures}

To assess whether temperature effects could be learned implicitly, we trained global dimensionality-reduction models using the full set of 280 bond-environment descriptors as inputs, combining data from all isothermal conditions. Both the feedforward autoencoder and the non-negative matrix factorization (NMF) approaches were applied in this global setting.

\begin{figure}[H]
    \centering
    \includegraphics[width=0.8\textwidth]{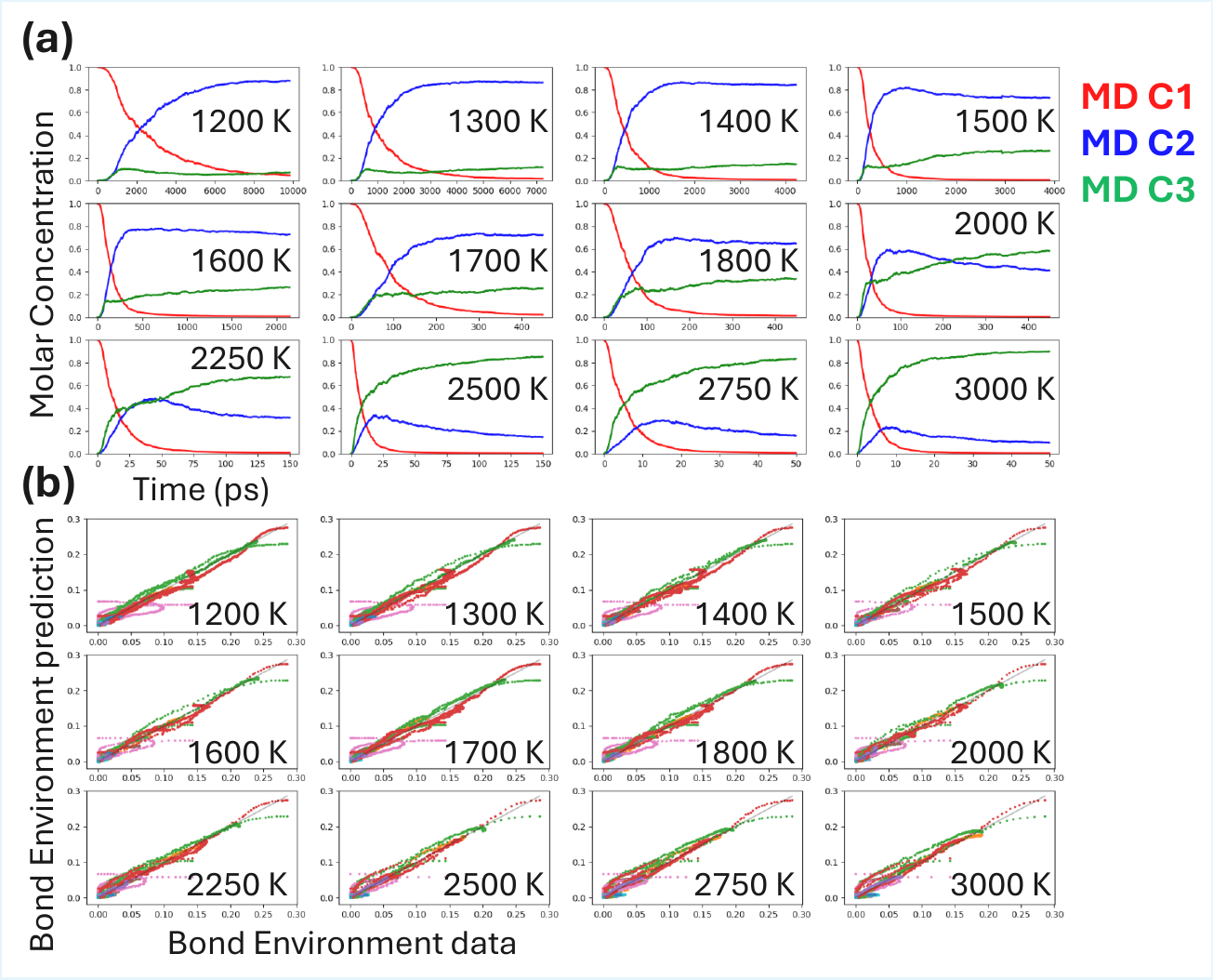}
    \caption{\label{FigureS1}Single feedforward autoencoder trained on bond-environment data from all isothermal temperatures simultaneously. (a) Three-component encoded representation, illustrating a temperature-dependent trade-off between the second and third components rather than the expected reactant-intermediate-product cascade. (b) Parity plot comparing decoded bond environments to molecular dynamics data, yielding an overall mean squared error of 4.97 × 10$^{-6}$.}
\end{figure}

\begin{figure}[H]
    \centering
    \includegraphics[width=0.8\textwidth]{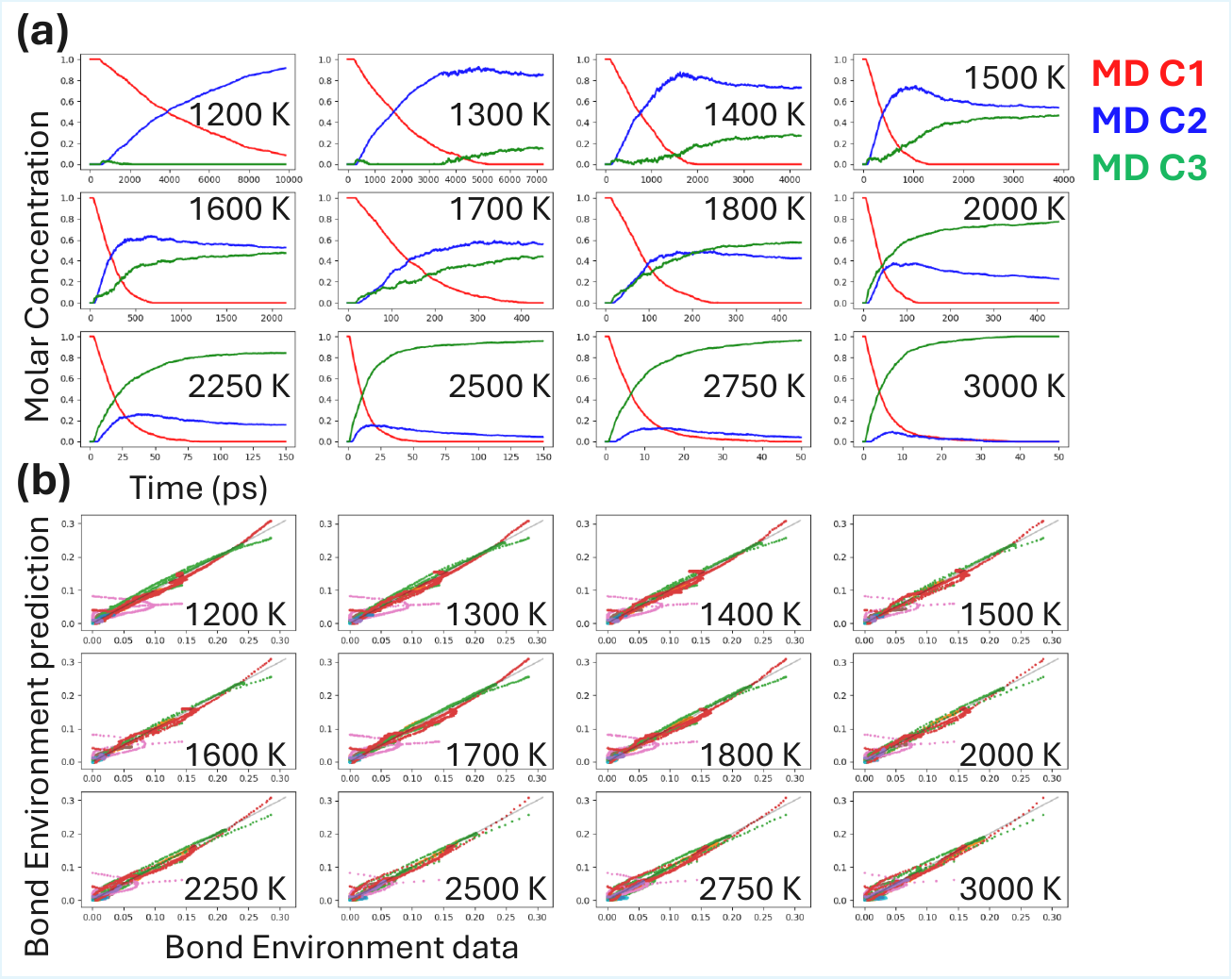}
    \caption{\label{FigureS2}Single non-negative matrix factorization (NMF) model trained on bond-environment data from all isothermal temperatures simultaneously. (a) Three-component decomposition exhibiting behavior similar to the global autoencoder, with a temperature-dependent trade-off between the second and third components rather than a clear reactant–intermediate–product progression. (b) Parity plot comparing reconstructed bond environments to molecular dynamics data, yielding an overall mean squared error of 4.19 × 10$^{-6}$, which is slightly lower than that of the corresponding autoencoder model.}
\end{figure}

Figures~\ref{FigureS1} and~\ref{FigureS2} present the resulting encoded components [panel (a)] and the corresponding parity plots [panel (b)] comparing reconstructed bond environments with the molecular dynamics data. In both cases, although the reconstruction errors are relatively low, the latent variables lack clear chemical interpretability. In particular, the three-component representations do not exhibit the expected sequential behavior associated with effective reactant, intermediate, and product populations. Instead, the encoded components display temperature-dependent trade-offs and mixed features that obscure their physical meaning.

These results indicate that, despite their ability to reduce reconstruction error, global models that attempt to infer temperature effects implicitly struggle to produce chemically intuitive latent spaces when the underlying decomposition pathways vary significantly with temperature. This observation motivates the development of the parametric, temperature-dependent autoencoder framework presented in the main text.

\section{Temperature-dependent encoder and decoder weights}

\begin{figure}[H]
    \centering
    \includegraphics[width=0.8\textwidth]{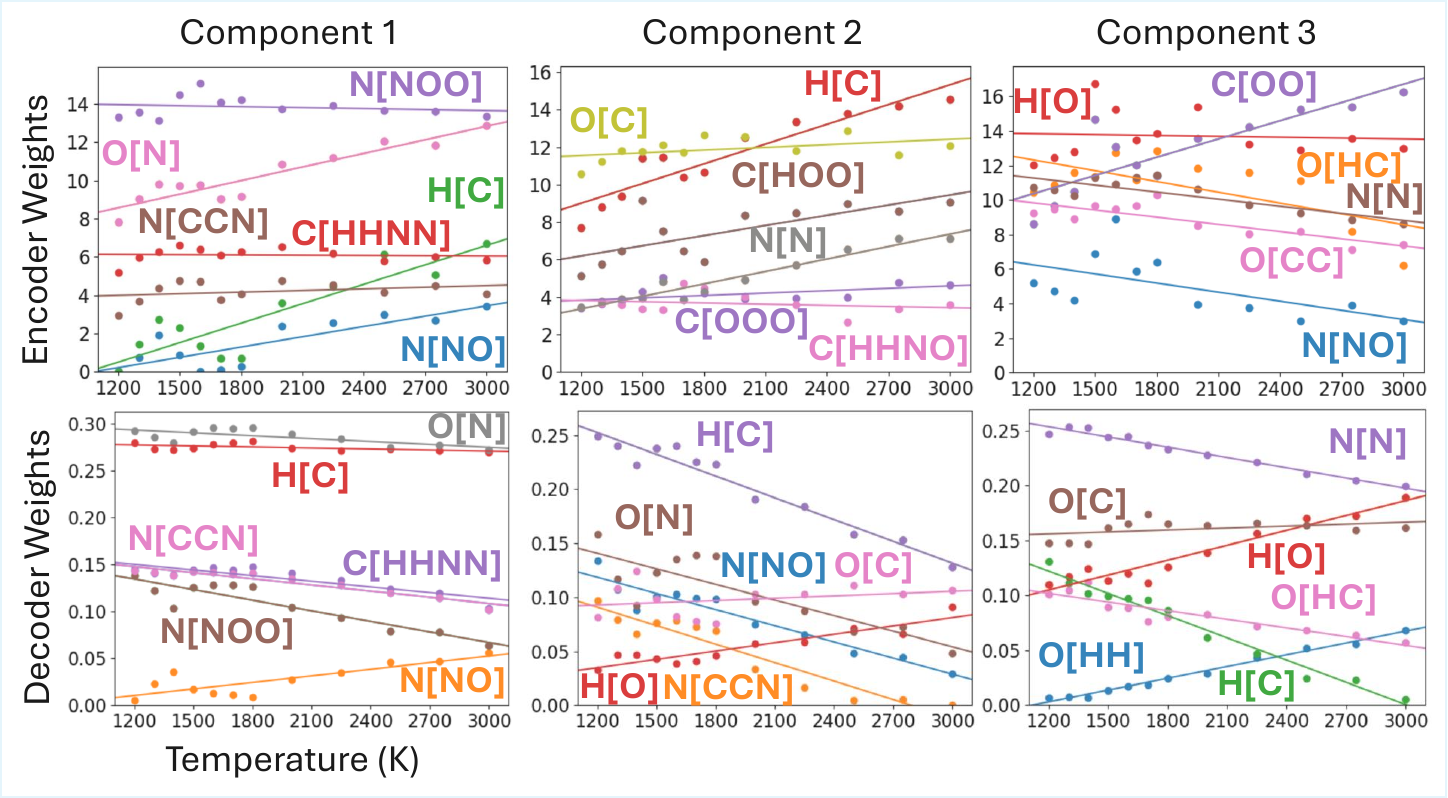}
    \caption{\label{FigureS3}Temperature dependence of selected bond-environment weights extracted from the encoder and decoder of the individual isothermal autoencoder models. Three bond environments were chosen based on having the largest absolute slopes with respect to temperature, while the remaining three were selected due to their large overall weight magnitudes. The smooth variation with temperature motivates the use of linear functional forms in constructing the parametric, temperature-dependent autoencoder.}
\end{figure}

From the set of autoencoder models trained independently at each isothermal temperature, we extracted the kernel weights associated with both the encoder and decoder layers for each latent component. Figure~\ref{FigureS3} illustrates representative examples of the six bond-environment descriptors with the largest magnitudes in the weight matrices for each component. These dominant environments identify the molecular motifs that contribute most strongly to the corresponding latent variable, either during encoding or reconstruction.

Analysis of these weights provides chemical insight into the decomposition process. In particular, the dominant environments correspond to molecular fragments and products expected to appear at different stages of decomposition, and their relative importance varies systematically with temperature. This behavior reflects the underlying thermally activated reaction pathways and the shifting balance between reactants, intermediates, and products as the thermodynamic conditions change.

Importantly, we observe that the temperature dependence of the encoder and decoder weights is smooth and well behaved across the studied range. Linear fits to the weight values as functions of isothermal temperature are also included in Fig.~\ref{FigureS3}. The resulting slopes and intercepts are subsequently extracted and used to parameterize the temperature-dependent weight matrices in the global, interpretable autoencoder described in the main text.

\section{Temperature-dependent non-negative matrix factorization model}

The construction of a temperature-dependent non-negative matrix factorization (NMF) model differs in important ways from the parametric, interpretable autoencoder proposed in the main text. In NMF, the data matrix is decomposed into the product of two lower-rank non-negative matrices, which correspond to an encoding matrix and a basis (weight) matrix. Their product reconstructs the original bond-environment data, analogous to the encoder–decoder structure of the autoencoder.

To introduce temperature dependence, we extracted the NMF basis matrices obtained from models trained independently at each isothermal temperature and fit temperature-dependent functions to the corresponding weights for each latent component. Using these fitted weights, we constructed temperature-specific basis matrices and computed their Moore–Penrose pseudoinverses.~\cite{strang1980} The encoded coefficients at each temperature were then obtained by multiplying the pseudoinverse of the temperature-dependent basis matrix by the original bond-environment data.

While this procedure yields a mathematically consistent reconstruction, the resulting latent variables are considerably less interpretable than those produced by the parametric autoencoder. In particular, the encoded components exhibit nonphysical behavior, including negative values and non-monotonic transients, especially at early times. These artifacts undermine the interpretation of the latent variables as effective chemical components, despite the relatively low reconstruction error. This comparison highlights the advantage of the parametric autoencoder framework, which enforces interpretability through architectural constraints while retaining comparable reconstruction accuracy.

\begin{figure}[H]
    \centering
    \includegraphics[width=0.8\textwidth]{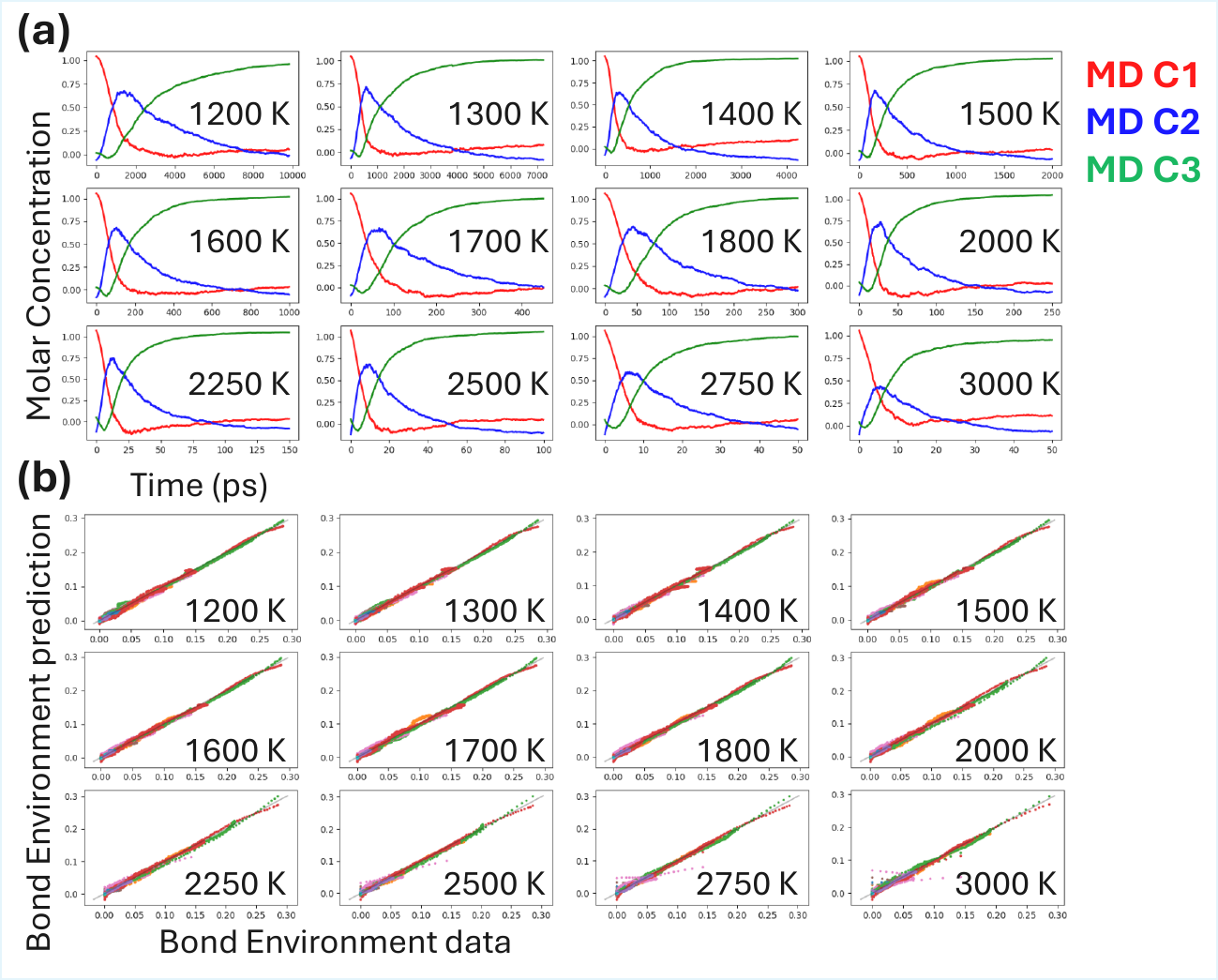}
    \caption{\label{FigureS4}Non-negative matrix factorization (NMF) model with explicitly temperature-dependent weights trained on bond-environment data from all isothermal simulations. (a) Three-component decomposition exhibiting behavior qualitatively similar to a first-order kinetic cascade, but with nonphysical features such as negative component values and an initial decrease in the third component (green). (b) Parity plot comparing reconstructed bond environments to molecular dynamics data, yielding an overall mean squared error of 1.52 × 10$^{-6}$, representing approximately a half-order-of-magnitude improvement over the corresponding global NMF and autoencoder models.}
\end{figure}

\section{Kinetic parameterization using non-negative matrix factorization models}

To benchmark the kinetic parameters obtained from the parametric, interpretable autoencoder, we applied the same kinetic fitting procedure to latent variables derived from two alternative non-negative matrix factorization (NMF) approaches: (i) a single global NMF model trained across all isothermal temperatures and (ii) an NMF model with explicitly temperature-dependent basis weights.
For each case, the reduced-order chemical kinetics model was fit to the corresponding three-component representations obtained from NMF. The resulting kinetic prefactors, activation energies, and associated mean squared errors (MSEs) are summarized below. This comparison highlights the impact of latent-space interpretability on the stability and accuracy of the fitted kinetic parameters, and provides a quantitative contrast with the results obtained using the parametric, interpretable autoencoder.

\begin{figure}[H]
    \centering
    \includegraphics[width=0.8\textwidth]{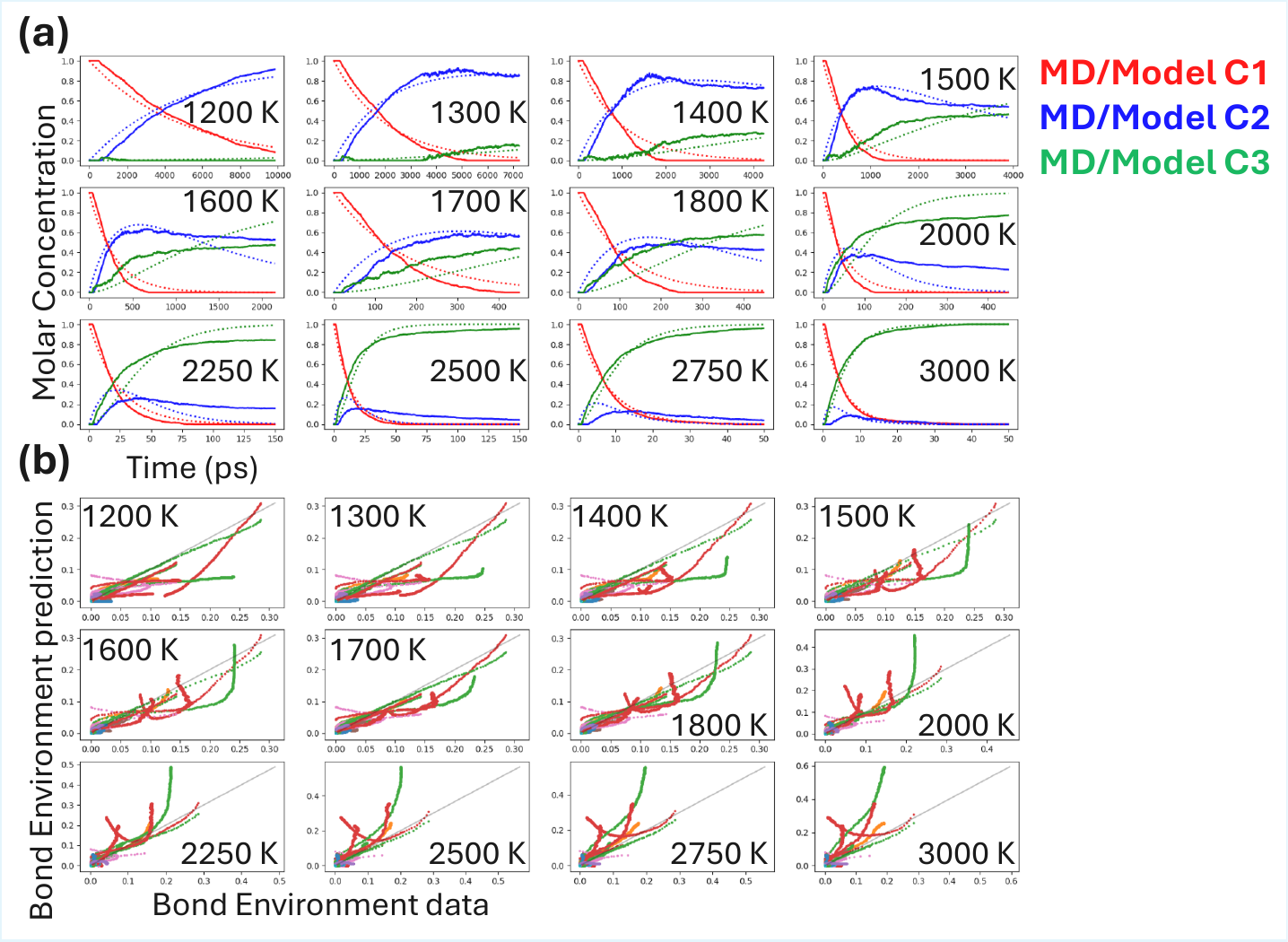}
    \caption{\label{FigureS5}Kinetic parameterization based on the reduced-order chemistry model derived from the global NMF decomposition shown in Fig.~\ref{FigureS2}. The reaction scheme described in the main text was assumed, and kinetic parameters were obtained by minimizing the mean squared error between the model-predicted and encoded component concentrations across all isothermal temperatures. (a) Best-fit kinetic parameters are $Z_1$ = 17.9 ps$^{-1}$, $E_1$ = 27.1 kcal/mol, $Z_2$ = 1811.9 ps$^{-1}$, and $E_2$ = 47.1 kcal/mol, yielding an overall concentration MSE of 7.25 × 10$^{-3}$. (b) Parity plot comparing decoded bond environments to molecular dynamics data, with a total MSE of 2.63 × 10$^{-4}$, indicating poor reconstruction of the latent space, particularly at intermediate temperatures.}
\end{figure}

\begin{figure}[H]
    \centering
    \includegraphics[width=0.8\textwidth]{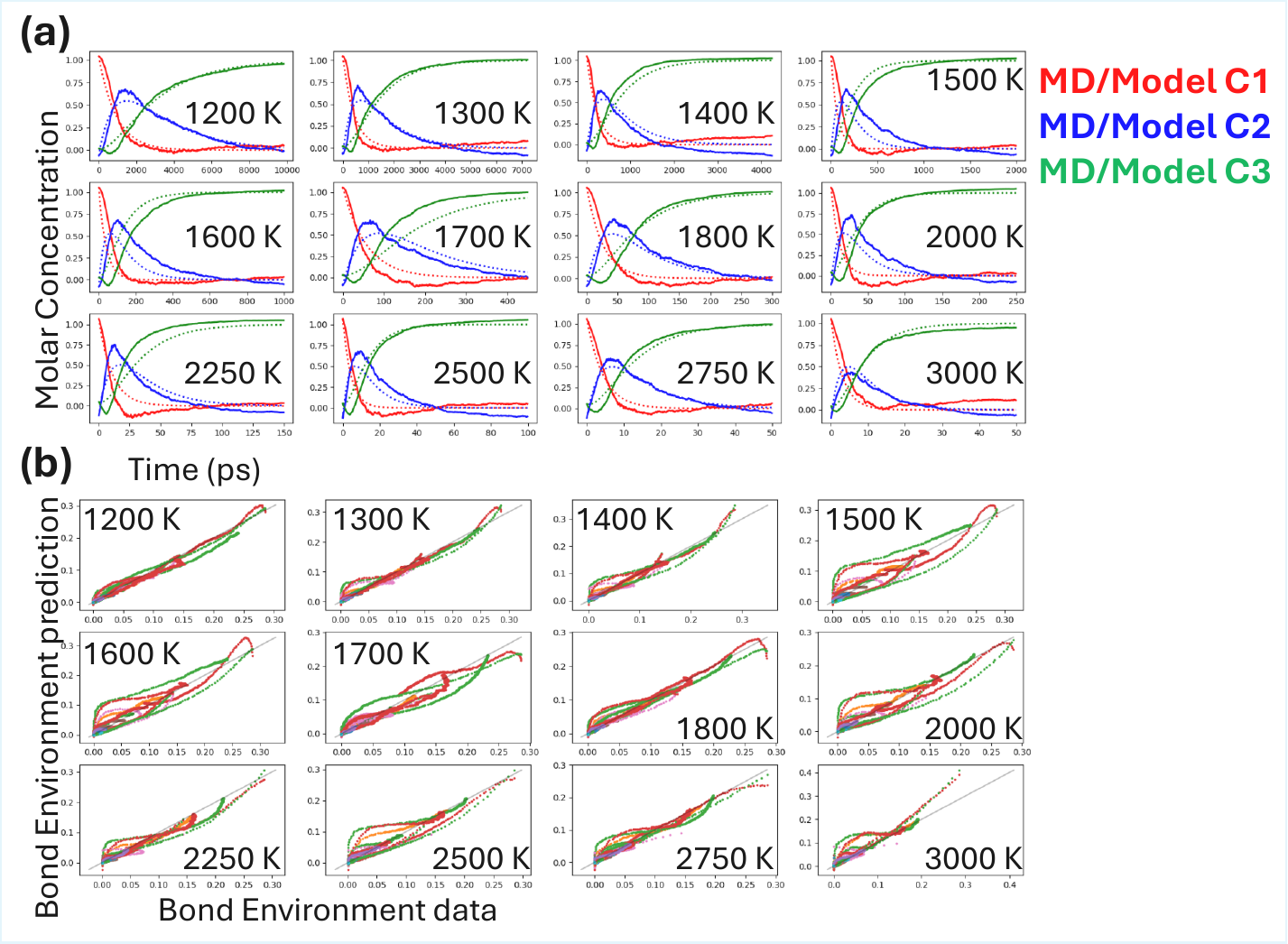}
    \caption{\label{FigureS6}Kinetic parameterization based on the non-negative matrix factorization (NMF) model with temperature-dependent weights, constructed following the procedure described in Sec. II.C. The reaction scheme outlined in the main text was assumed, and kinetic parameters were obtained by minimizing the mean squared error between the model-predicted and encoded component concentrations across all isothermal temperatures. (a) Best-fit parameters are $Z_1$ = 14.1 ps$^{-1}$, $E_1$ = 22.6 kcal/mol, $Z_2$ = 7.5 ps$^{-1}$, and $E_2$ = 23.4 kcal/mol, yielding an overall concentration MSE of 7.29 × 10$^{-3}$. (b) Parity plot comparing decoded bond environments to molecular dynamics data, with a total MSE of 6.90 × 10$^{-6}$, demonstrating substantially improved reconstruction of the latent space relative to the global NMF model shown in Fig.~\ref{FigureS5}.}
\end{figure}

\section{Prediction of an unseen adiabatic temperature trajectory}

To assess the predictive capability of the proposed framework, we validated the model against an adiabatic molecular dynamics simulation conducted at an initial temperature not included in the training set. The reduced-order chemical kinetics parameters were taken directly from the fits to the isothermal parametric, interpretable autoencoder components, while the heat-release parameters were obtained from the adiabatic temperature profiles described in the main text.

Using these independently determined parameters, the model was used to predict the time evolution of component concentrations and temperature under adiabatic conditions. Comparisons with the molecular dynamics results demonstrate good agreement in both the concentration dynamics and temperature rise. In addition, parity plots of the decoded bond-environment populations show close correspondence with the underlying atomistic data. These results confirm that the parametric autoencoder–based reduced-order model generalizes well to previously unseen thermodynamic conditions.

\begin{figure}[H]
    \centering
    \includegraphics[width=\textwidth]{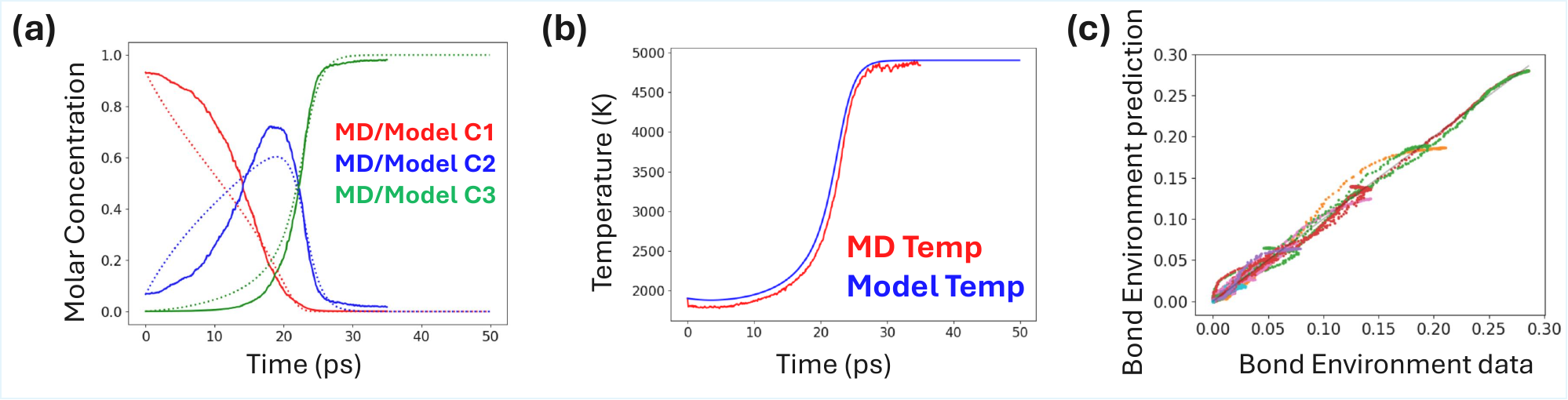}
    \caption{\label{FigureS7}Validation of the parametric, interpretable autoencoder using an adiabatic simulation at an unseen initial temperature $T_0$ = 1900 K. (a) Time evolution of the encoded component concentrations and (b) temperature evolution, comparing molecular dynamics results with model predictions, show good agreement, with a concentration mean squared error of 7.67 × 10$^{-3}$ and a temperature mean absolute error of 11.7 K. (c) Parity plot comparing decoded bond environments to molecular dynamics data, yielding a mean squared error of 2.14 × 10$^{-6}$, which is of the same order of magnitude as the overall training error reported in Fig. 3.}
\end{figure}

\section{Training reduced-order chemical kinetics and heat-release parameters using Keras}

To complement the standalone parameter optimization described in the main text, we implemented one-step propagating neural network models in Keras to optimize the reduced-order chemical kinetics parameters from isothermal simulations and the heat-release parameters from adiabatic simulations. These networks are constructed to mirror the underlying physical equations governing concentration and temperature evolution, while treating the kinetic prefactors, activation energies, and heat-release constants as trainable parameters.

For the isothermal case, the network takes the component concentrations at time $t$, along with the temperature and timestep, as inputs and predicts the concentrations at time $t+\Delta t$ using a single-step propagation consistent with the kinetic rate equations. Similarly, for the adiabatic case, the network augments the concentration update with a temperature evolution step that incorporates the heat-release formulation. Training is performed by minimizing the mean squared error between the predicted and reference trajectories obtained from molecular dynamics simulations.

An example training workflow is provided below, illustrating convergence of the neural network–based optimization to the same kinetic and heat-release parameters obtained from the standalone numerical fitting procedures. This agreement demonstrates that the Keras implementation faithfully reproduces the physics-based reduced-order model and provides a flexible framework for future extensions, including joint optimization with the parametric autoencoder in an end-to-end learning setting.

\begin{figure}[H]
    \centering
    \includegraphics[width=\textwidth]{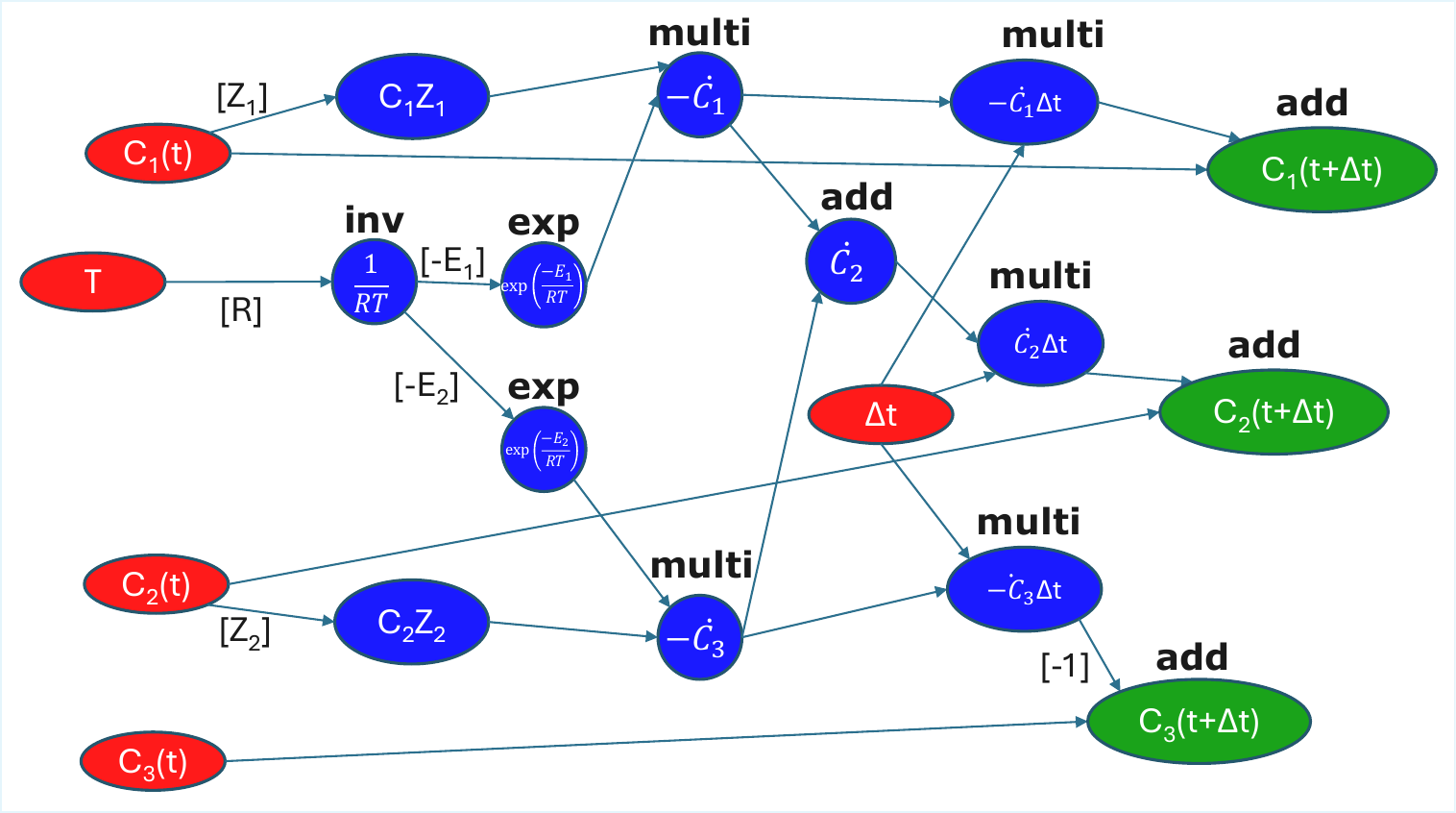}
    \caption{\label{FigureS8}Neural-network implementation of the reduced-order chemical kinetics model in Keras. Nodes are colored by role: inputs (red), hidden layers (blue), and outputs (green). Weights are indicated explicitly in brackets where applicable; otherwise, they are assumed to be unity. The only trainable parameters in the network are the kinetic prefactors and activation energies, $Z_1$, $E_1$, $Z_2$, and $E_2$. Activation functions are labeled above the corresponding nodes, with unlabeled nodes assumed to use linear activations.}
\end{figure}

\begin{figure}[H]
    \centering
    \includegraphics[width=0.8\textwidth]{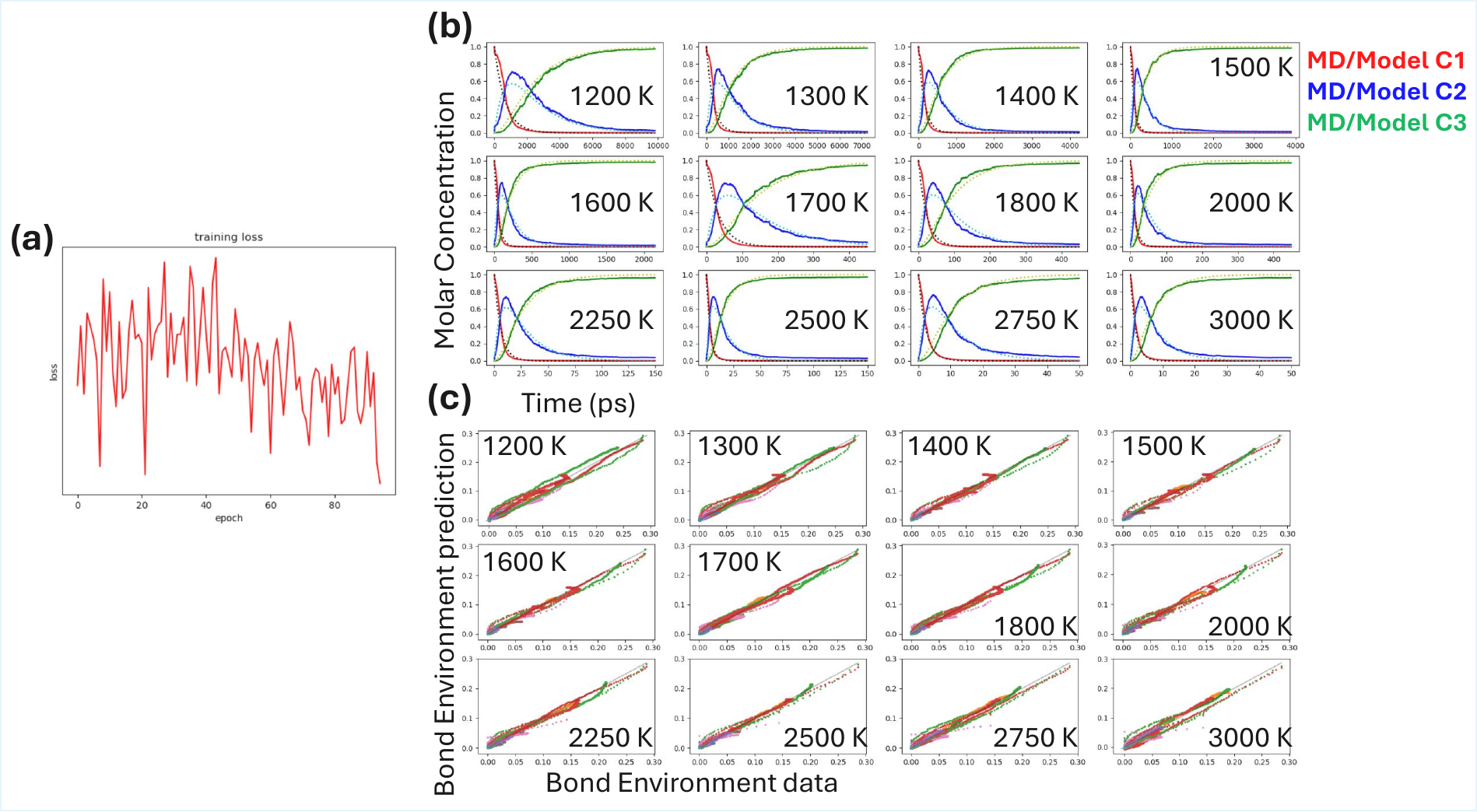}
    \caption{\label{FigureS9}Training of the reduced-order chemical kinetics model from Fig. 4 implemented in Keras. The reaction scheme described in the main text was assumed. (a) Mean squared error (MSE) as a function of training epoch shows no further decrease, indicating that the manually optimized kinetic parameters already correspond to an optimal solution. (b) The learned kinetic parameters and concentration MSE remain unchanged relative to those reported in Fig. 4. (c) Parity plot comparing decoded bond environments to molecular dynamics data shows no improvement in MSE compared to Fig. 4, confirming consistency between the stand-alone and neural-network implementations.}
\end{figure}

\begin{figure}[H]
    \centering
    \includegraphics[width=\textwidth]{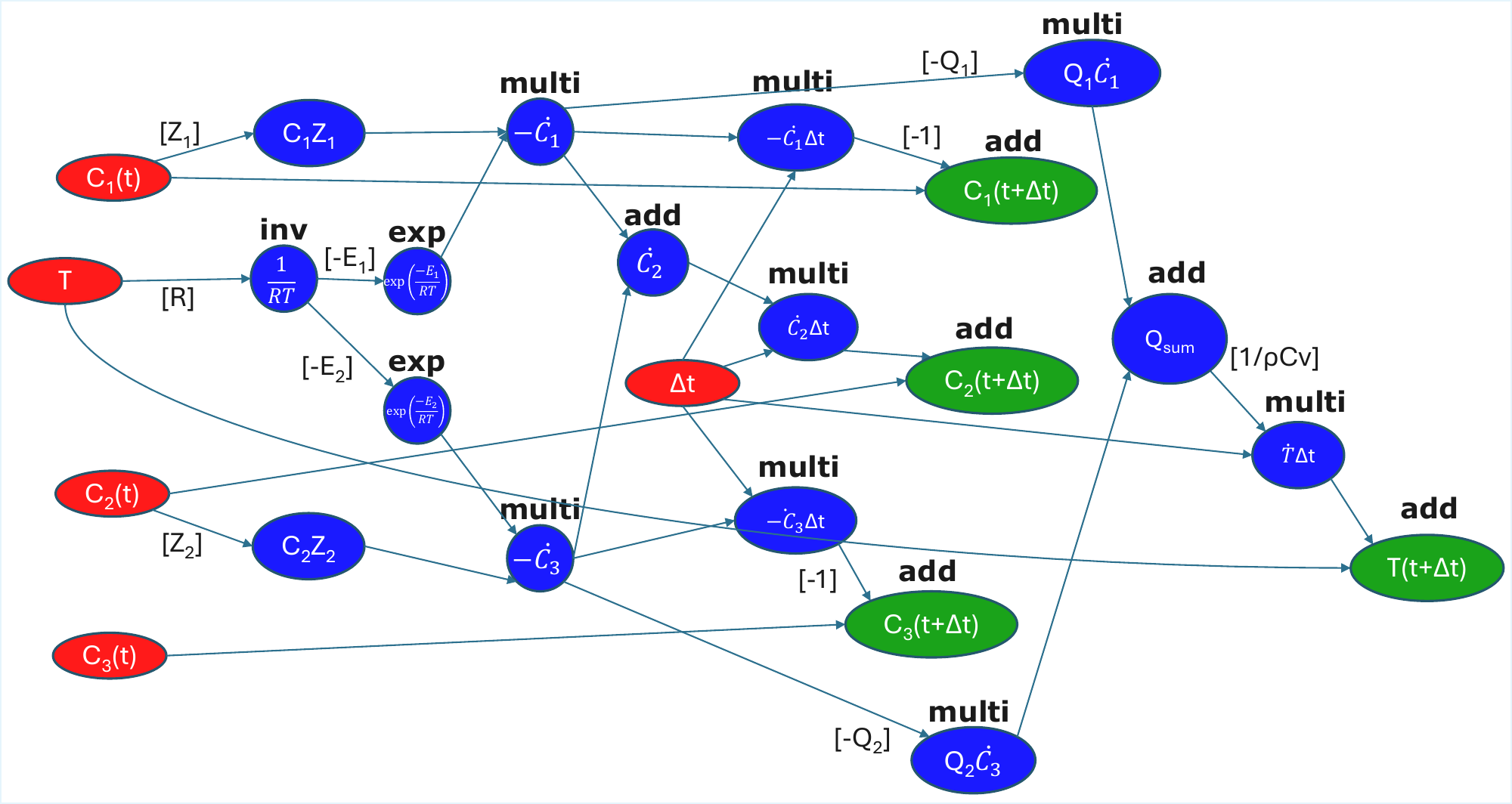}
    \caption{\label{FigureS10}Neural-network implementation of the reduced-order chemical kinetics model with heat release in Keras. The architecture extends the isothermal model shown in Fig.~\ref{FigureS8} by incorporating the concentration time derivatives into an adiabatic temperature update. In contrast to the isothermal single-step prediction network, temperature is treated as an explicit output variable, enabling the model to capture thermo–chemical feedback associated with adiabatic heating.}
\end{figure}